\documentclass{aa}  
\usepackage{natbib}
\bibpunct{(}{)}{;}{a}{}{,}
\usepackage{graphicx}
\usepackage{txfonts}
\usepackage{xcolor}
\usepackage{textcomp} 
\usepackage[version=3]{mhchem} 
\usepackage[squaren]{SIunits}
\usepackage{dcolumn}
\usepackage{longtable}
\usepackage{supertabular}
\usepackage{booktabs} 

\newcommand{\rev}[1]{#1}

\def\decsec {\ifmmode {\rlap.}$\,$^{\rm s}$\,$\! \else ${\rlap.}$\,$^{\rm s}$\,$\!$\fi}\def\decs {\ifmmode {\rlap.}$\,$^{\rm s}$\,$\! \else ${\rlap.}$\,$^{\rm
s}$\,$\!$\fi}

\begin{document}

\title{Submillimeter spectroscopy and astronomical searches of vinyl mercaptan, C$_2$H$_3$SH}

\author{
	M.-A. Martin-Drumel\inst{1}\fnmsep\inst{2}\fnmsep\thanks{Present address: Institut des Sciences Mol\'eculaires d'Orsay (ISMO), CNRS, Univ. Paris-Sud, Universit\'e Paris-Saclay, F-91405 Orsay, France}
    \and
	K. L. K. Lee \inst{2}
    \and
	A. Belloche\inst{3}
    \and
	O. Zingsheim\inst{1}
	\and
	S. Thorwirth\inst{1}
	\and
	H. S. P. M\"uller\inst{1}
	\and
	F. Lewen\inst{1}
	\and
	R. T. Garrod\inst{4}
	\and
	K. M. Menten\inst{3}
	\and
	M. C. McCarthy\inst{2}
	\and
	S. Schlemmer\inst{1}
	}

\institute{
	I. Physikalisches Institut, Universit\"at zu K\"oln, Z\"ulpicher Str. 77, 50937 K\"oln, Germany\\
	\email{marie-aline.martin@u-psud.fr}
	\and
	Harvard-Smithsonian Center for Astrophysics, 60 Garden Street, Cambridge, MA 02138, United States
	\and
	Max-Planck-Institut f\"ur Radioastronomie, Auf dem H\"ugel 69, 53121 Bonn, Germany
	\and
	Departments of Astronomy \& Chemistry, University of Virginia, McCormick Road, PO Box 400319, Charlottesville, VA 22904-4319, United States
	}

\date{}

 
  \abstract
   {New laboratory investigations of the rotational spectrum of postulated astronomical species are essential to support the assignment and analysis of current astronomical surveys. In particular, considerable interest surrounds sulfur analogs of oxygen-containing interstellar molecules and their isomers.}
   {To enable reliable interstellar searches of vinyl mercaptan, the sulfur-containing analog to the astronomical species vinyl alcohol, we have investigated its pure rotational spectrum at millimeter wavelengths.}
   {We have extended the pure rotational investigation of the two isomers \textit{syn} and \textit{anti} vinyl mercaptan to the millimeter domain using a frequency-multiplication spectrometer. The species were produced by a radiofrequency discharge in 1,2-ethanedithiol. Additional transitions have been re-measured in the centimeter band using Fourier-transform microwave spectroscopy to better determine rest frequencies of transitions with low-$J$ and low-$K_a$ values. Experimental investigations were supported by quantum chemical calculations on the energetics of both the [C$_2$,H$_4$,S] and [C$_2$,H$_4$,O] isomeric families. Interstellar searches for both \textit{syn} and \textit{anti} vinyl mercaptan as well as vinyl alcohol were performed in the EMoCA \rev{(Exploring Molecular Complexity with ALMA)} spectral line survey carried out toward \rev{Sagittarius (Sgr) B2(N2)} with \rev{the Atacama Large Millimeter/submillimeter Array (ALMA)}.}
   {Highly accurate experimental frequencies (to better than 100 kHz accuracy) for both \textit{syn} and \textit{anti} isomers of vinyl mercaptan have been measured up to 250 GHz; these  deviate considerably from predictions based on extrapolation of previous microwave measurements. Reliable frequency predictions of the astronomically most interesting millimeter-wave lines for these two species can now be derived from the best-fit spectroscopic constants. From the energetic investigations, the four lowest singlet isomers of the [C$_2$,H$_4$,S] family are calculated to be nearly isoenergetic, which makes this family a fairly unique test bed for assessing possible reaction pathways. Upper limits for the column density of \textit{syn} and \textit{anti} vinyl mercaptan are derived toward the extremely molecule-rich star-forming region Sgr B2(N2) enabling comparison with selected complex organic molecules.}
   {}


   \maketitle
%

\section{Introduction} 

From the diatomic \ce{SH+} and SH radicals \citep{Menten2011,Neufeld2012} to the large complex organic molecule ethyl mercaptan (ethanethiol, \ce{C2H5SH}, \citeauthor{Kolesnikova2014} \citeyear{Kolesnikova2014}),
nearly 20 sulfur-bearing species have been observed in the interstellar medium (ISM) and in the circumstellar envelopes of evolved stars.  This total is substantial, representing roughly 10\,\% of all the known interstellar or circumstellar compounds.   While the abundance of S-containing species in the diffuse ISM is consistent with the estimated cosmic abundances of sulfur \citep{Savage1996}, and is well reproduced by current astrophysical models \citep{Neufeld2015}, the same set of compounds only accounts for $\sim$0.1\,\% of the expected sulfur abundance in the cold, dense clouds and circumstellar regions around young stellar objects \citep{Joseph1986, Tieftrunk1994}. This remarkable disparity has given rise to a simple, but currently unanswered, question: where is the missing sulfur? Several potential reservoirs have been investigated, including  atomic sulfur \citep{Anderson2013}, S-bearing species trapped on icy dust grains \citep{Jimenez-Escobar2011}, and S-containing molecules in the gas phase \citep{Martin-Domenech2016, Cazzoli2016}, along with other possibilities. From studies of shocked gas, however, atomic sulfur can only account for at most 5--10\,\% of the total sulfur cosmic abundance \citep{Anderson2013}, implying that additional sulfur sinks must exist. Many studies propose that the depletion of sulfur reflects accretion onto dust grains, where it could be sequestered in various forms as inorganic and organic molecules \citep{Scappini2003}, or atomic and polymeric S$_n$ chains and rings \citep{Wakelam2004}. So far, however, only two sulfur molecules, OCS and \ce{SO2}, have been detected on grains, both in too low abundance for either species to be a major reservoir of sulfur \citep{Palumbo1995, Boogert1997}. 
Since \ce{H2S} is the most abundant sulfur-bearing molecule in cometary ices \citep{Bockelee-Morvan2000}, its presence on ice grains has also received some attention, but the rather low upper limit on its abundance also appears to rule out this species as a major sink for sulfur \citep{Smith1991, Jimenez-Escobar2011}. 
Finally, photoprocessing of sulfur-containing ice products could yield gaseous S-bearing species in the dense ISM in molecular forms that are yet to be identified \citep{Jimenez-Escobar2011}. Consequently, the identification of missing interstellar sulfur carriers requires the detection of new gaseous sulfur
containing compounds, which in turn requires laboratory investigations of their rotational spectra. Transient species are particularly appealing candidates, but are often challenging to characterize in the laboratory compared to their stable, often commercially available, counterparts, especially in the millimeter and submillimeter wave region where high-altitude, high-sensitivity radio facilities such as \rev{the Atacama Large Millimeter/submillimeter Array (ALMA)} operate.

While \cite{Kolesnikova2014} reported the detection of ethyl mercaptan in Orion, its presence there is not without controversy \citep{Mueller16}. Regardless, experimental data on related molecules that might also be present in the ISM and participate in the same chemical network, for instance unsaturated variants, may prove helpful in resolving this controversy. The isomeric family with the elemental formula [C$_2$,H$_4$,S] --- thioacetaldehyde (\ce{CH3CHS}), thiirane (\textit{c}-\ce{C2H4S}), and vinyl mercaptan (\ce{CH2CHSH}) --- are of particular  interest as all three isovalent oxygen counterparts (acetaldehyde, oxirane, and vinyl alcohol) have been detected in the ISM \citep{Gottlieb1973, Dickens1997, Turner2001}. 
Structural isomers are an invaluable tool to probe the chemical dynamics of astrophysical environments. While a minimum energy principle was once proposed to assert molecular abundance in space \citep{Lattelais2009}, the general consensus is that chemical kinetics, rather than thermodynamic stability, is driving the formation of molecules in astronomical sources \citep{Loomis2015}. 
Interestingly, relatively little is known about isomers of relatively large astrophysical species (5 atoms and more), in part due to the difficulties involved in both producing and detecting these species in the laboratory, and the [C$_2$,H$_4$,S] family is no exception. While the pure rotational spectrum of each of these three singlet isomers has been investigated in the microwave domain \citep{Kroto1974, CunninghamJr.1951, Tanimoto1977}, high-resolution  measurements at millimeter and shorter wavelengths have only been published for thiirane \citep{Hirao2001, Bane2012}, a likely reflection of the stability of this compound in the laboratory. In contrast, thioacetaldehyde polymerizes rapidly, with a lifetime of the order of ten seconds \citep{Kroto1974}, and consequently it must, as with vinyl mercaptan, be produced \textit{in situ} for spectroscopic studies. To confidently detect these two transient species in the ISM, laboratory measurements at higher frequency are essential, particularly in spectral bands in which ALMA operates.

In this study the pure rotational spectrum of the \textit{syn} and \textit{anti} isomers of vinyl mercaptan in their ground vibrational state has been measured up to 250\,GHz. From the newly-recorded transitions, highly accurate rotational and centrifugal distortion constants have been derived for both species, ensuring that astronomical searches in the millimeter domain can now be undertaken with confidence. 
Using the new experimental data, we carried searches on the spectral line survey EMoCA (Exploring Molecular Complexity with ALMA) performed toward the high-mass star forming region \rev{Sagittarius (Sgr) B2(N)} \citep{Belloche16}.
As part of this work, we also investigated the energetics of the [\ce{C2},\ce{H4},S] isomeric family and its isovalent oxygen counterparts by high-level quantum chemical calculations.

\section{Methods}

\subsection{Spectroscopic background}

\begin{figure}[ht!]
\centering
\includegraphics[width=0.45\textwidth]{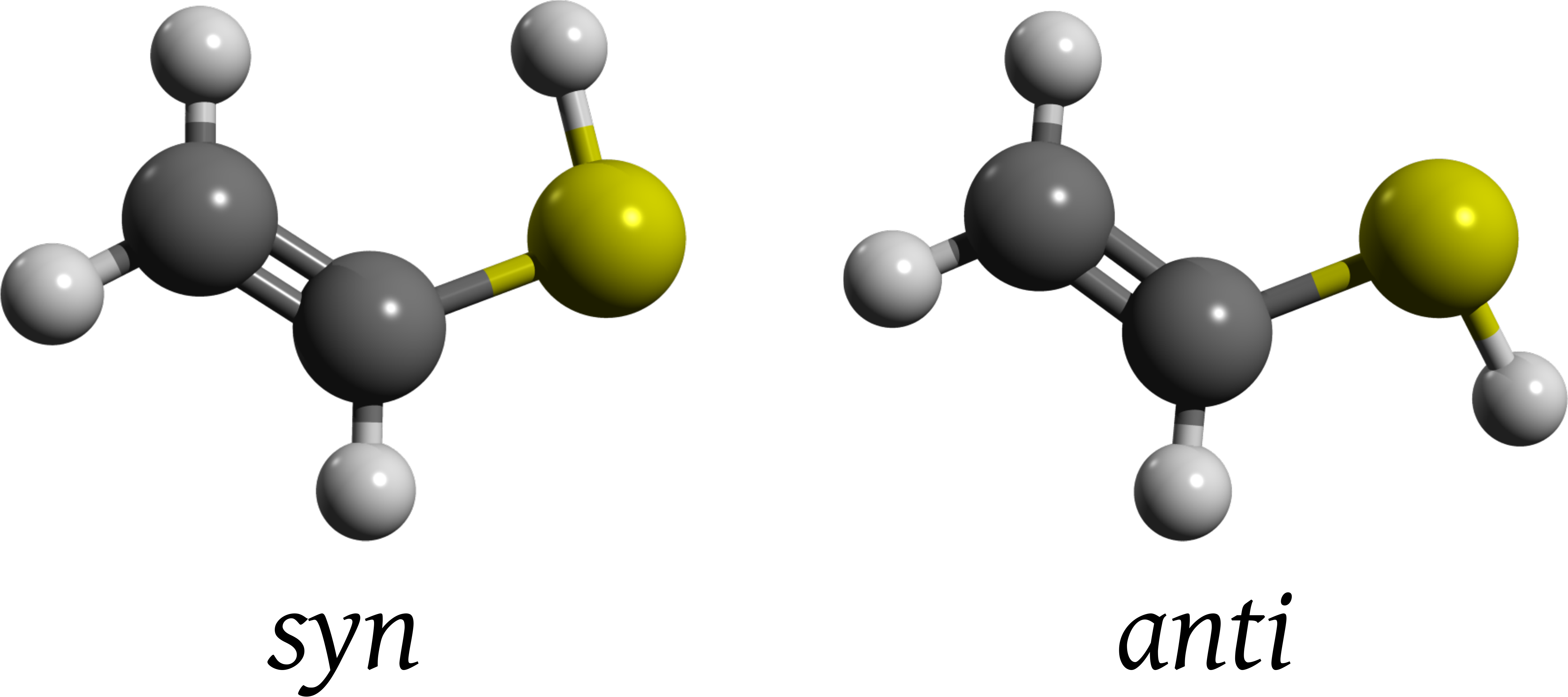}
\caption{Molecular representation of planar \textit{syn} and quasi-planar \textit{anti} vinyl mercaptan.}
\label{fig: molecules}
\end{figure}

Vinyl mercaptan exists in the gas phase in two distinct rotameric forms, the planar \textit{syn} and quasi-planar \textit{anti} isomers (Fig. \ref{fig: molecules}), with the former only slightly more stable than the latter by about 0.6\,kJ/mol \citep{Almond1983, Almond1985, Plant1992}, which yields a \textit{syn/anti} ratio at room temperature  of about 1.2. 
It is worth noting that while the equilibrium structure of the \textit{anti} isomer is non-planar, with a calculated SH  torsional angle of about 130--150\,\degree, the barrier to linearity (0.14\,kJ/mol) is small compared to the zero-point vibration, which  results in a vibrationally-averaged planar structure \citep{Tanimoto1979a, Almond1985}. As a consequence, both isomers can be considered as belonging to the $C_s$ point group symmetry \citep{Almond1983}. 

Experimental studies on vinyl mercaptan are limited. Photoelectron \citep{Chin1994} and low-resolution vibrational spectra \citep{Almond1977,Almond1983} have been reported, while pure rotational spectra of the two isomers in the ground and several excited vibrational states have only been measured up to 50\,GHz \citep{Almond1977, Tanimoto1977, Tanimoto1979, Tanimoto1979a}. Isotopic investigations of the microwave spectra of the SD isotopologues\footnote{i.e., in which the single hydrogen atom bonded to the S atom is replaced by a deuterium atom.} \citep{Tanimoto1977, Tanimoto1979, Tanimoto1979a}, and  extensive studies of deuterated variants of the \textit{syn} isomer \citep{Almond1985} that followed, enabled a determination of a partial geometry; owing to the absence of experimental data for isotopic species involving substitution at any of the three heavy atoms, a complete structure has not been reported. Additionally, accurate dipole moments in the ground vibrational state are available from Stark measurements: $\mu_a = 0.813 (1)$\,D, $\mu_b = 0.376 (4)$\,D, $\mu_{tot}=0.896 (3)$\,D for the \textit{syn} isomer; and $\mu_a = 0.425 (10)$~D, $\mu_b = 1.033 (10)$\,D, $\mu_{tot}=1.117 (14)$\,D for the \textit{anti} one \citep{Tanimoto1979, Tanimoto1979a}.

\subsection{Computational details}
Quantum-chemical calculations of the [\ce{C2},\ce{H4},S] and [\ce{C2},\ce{H4},O] isomeric families
were performed at the coupled-cluster level with single, double, and perturbative triple excitations [CCSD(T)] \citep{raghavachari_chemphyslett_157_479_1989}. All calculations were performed using the CFOUR \rev{(Coupled-Cluster techniques for Computational Chemistry)} program package \citep[][]{cfour-kelvin,harding_JChemTheoryComput_4_64_2008} in combination with Dunning's hierarchies of correlation consistent polarized valence and polarized core-valence sets \citep[][]{dunning_JCP_114_9244_2001,peterson_JCP_117_10548_2002}.
The best estimate equilibrium structures have been calculated at the all-electron (ae) CCSD(T)/cc-pwCVQZ level of theory, which has previously been shown to yield equilibrium structures of very high quality for molecules harboring second-row elements
\citep[see, e.g.,][]{coriani_JCP_123_184107_2005,thorwirth_JCP_2009,thorwirth_hccnsi_2015}. 
Centrifugal distortion constants and zero-point vibrational corrections $\Delta B_0$ to the rotational constants were calculated in the frozen-core (fc) approximation
at the fc-CCSD(T)/cc-pV(T+$d$)Z level and ground state rotational constants were obtained as $B_0=B_e-\Delta B_0$.
Here, the equilibrium rotational constants $B_e$ are calculated from the ae-CCSD(T)/cc-pwCVQZ structures. The calculated spectroscopic parameters for both \textit{syn} and \textit{anti} vinyl mercaptan are given in Table \ref{tab: param}, and the elements of their Z-matrices are given in Appendix \ref{ap:zmat}.

The relative energetics of the [\ce{C2},\ce{H4},O] and [\ce{C2},\ce{H4},S] isomers were calculated using a composite thermochemistry scheme similar to the HEAT \rev{(High Accuracy Extrapolated Ab initio Thermochemistry)} approaches \citep[see ][]{bomble_high-accuracy_2006,harding_high-accuracy_2008}. Using the CCSD(T)/cc-pwCVQZ geometries, a series of additive contributions are calculated: the extrapolated CCSD(T)/aug-cc-pCV$X$Z ($X$ = D, T, Q) correlation energy; the correlation energy from perturbative quadruple excitations with fc-CCSDT(Q)/cc-pVDZ; the anharmonic zero-point contribution with CCSD(T)/cc-pV(T+$d$)Z; the diagonal Born-Oppenheimer correction with HF/aug-cc-pVTZ; 1- and 2-electron scalar relativistic corrections with CCSD(T)/aug-cc-pCVTZ. 

\begin{figure}
    \centering
    \includegraphics[width=0.49\textwidth]{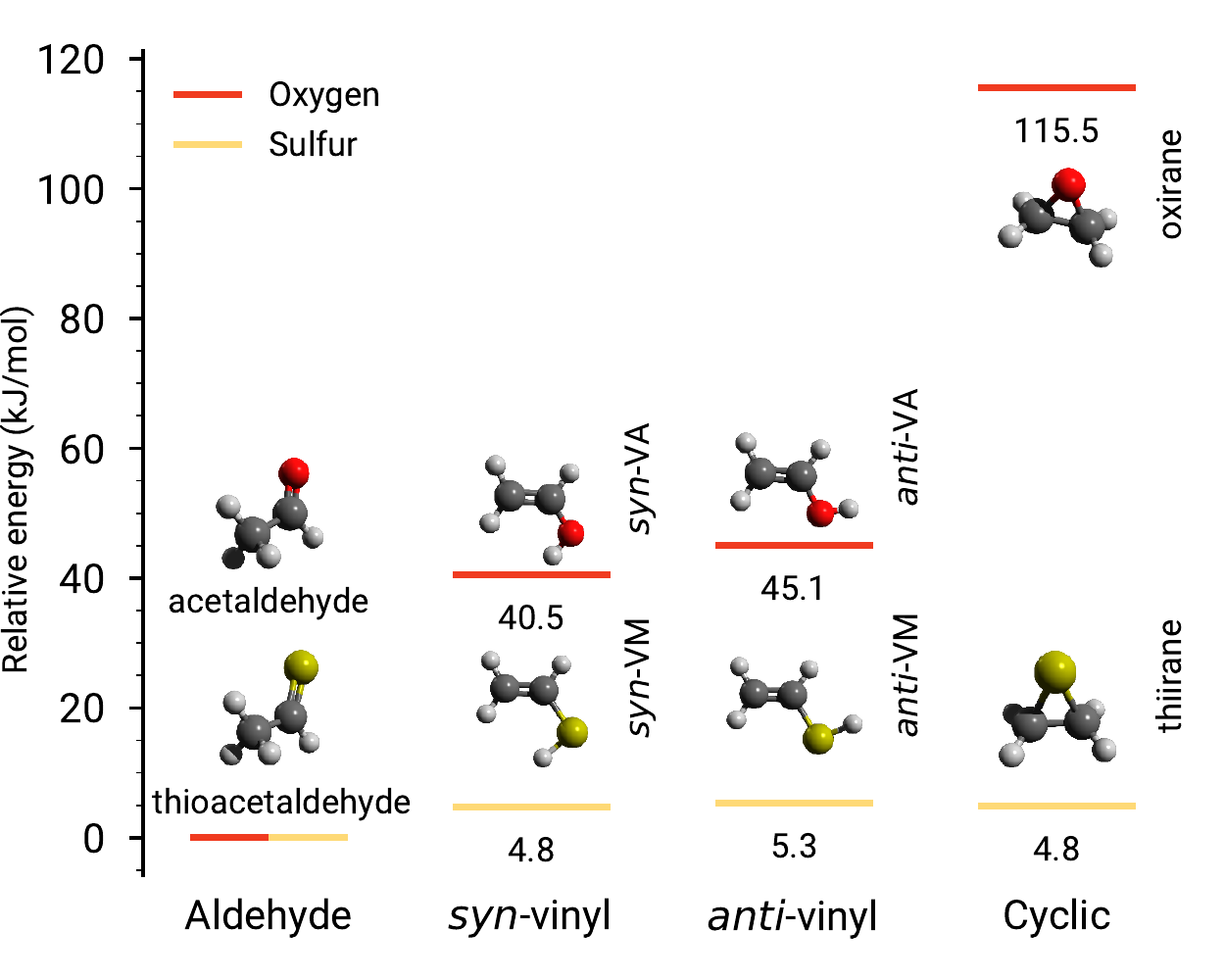}
    \caption{Energetics of the [\ce{C2},\ce{H4},$X$] species ($X$ = O or S), relative to the respective aldehydic forms, i.e. acetaldehyde and thioacetaldehyde, both set to 0 kJ/mol. } 
    \label{fig:relativeenergies}
\end{figure}

The relative stabilities -- with \ce{CH3CHO} and \ce{CH3CHS} as the point of reference -- are shown in Fig. \ref{fig:relativeenergies}. The energetic ordering of [\ce{C2},\ce{H4},O] isomers is in agreement with previous calculations by \citet{Lattelais2009} and \citet{karton_pinning_2014}: \ce{CH3CHO} is the most stable isomer, followed by vinyl alcohol and oxirane. In particular, the energetics computed at the W2-F12 level by \citet{karton_pinning_2014} for the \textit{syn} isomer of vinyl alcohol are within ${\sim}1$\,kJ/mol of the values presented here. 
The energy difference between the \textit{syn} and \textit{anti} isomers has been examined theoretically \citep{nobes_equilibrium_1981} and experimentally; the value presented here (4.6\,kJ/mol) is in excellent agreement with both the microwave ($4.5\pm0.6$, \cite{RODLER1985}) and the far-infrared semi-experimental determination (4.0\,kJ/mol, \cite{bunn_far-infrared_2017}). The sulfur analogues [\ce{C2},\ce{H4},S] on the other hand are near-isoenergetic at the current level of treatment: with the exception of \ce{CH3CHS}, the three remaining isomers are within ${\sim}0.5$\,kJ/mol of one another, with \textit{anti}-vinyl mercaptan being the highest in energy (5.3\,kJ/mol). Our \textit{ab initio} energy difference between the vinyl mercaptan isomers (0.5\,kJ/mol) is in quantitative agreement with the estimate by \cite{Almond1985} (0.6\,kJ/mol).

\subsection{Millimeter wave spectroscopy}

Millimeter wave measurements on vinyl mercaptan were performed using the Cologne (sub)-millimeter direct absorption spectrometer in the spectral regions 70--120\,GHz and 170--250\,GHz. The radiation was produced by a commercial amplifier-multiplier chain (\rev{Virginia Diodes Inc.}), and optimized by in-house electronics. To reduce $1/f$ noise, frequency modulation  was used with a  $2f$ demodulation, yielding lineshapes close to a second-derivative for all recorded transitions.
The 5 meter-long discharge cell, equipped with a roof-top mirror on one end, allows 10\,m of absorption path length.  
A detailed description of the apparatus can be found in \cite{Martin-Drumel2015}. 
Vinyl mercaptan was produced by a radio-frequency (RF) discharge in a \rev{continuous} flow of 1,2-ethanedithiol (\ce{HSCH2CH2SH}) at a pressure of about 10~\textmu bar. It should be noted, however, that since the pressure was measured between the cell and the pumping stage, a sizable pressure gradient may have been present in the cell.
Similarly to a previous study on OSSO \citep{Martin-Drumel2015}, the strongest signals  of vinyl mercaptan were observed when the RF power was reduced to fairly low powers. However, to maintain stable discharge conditions in the cell, and thus to limit the noise induced by a ``blinking'' plasma, a discharge power of no less than 5\,W was applied.

\begin{figure}[ht!]
\centering
\includegraphics[width=0.49\textwidth]{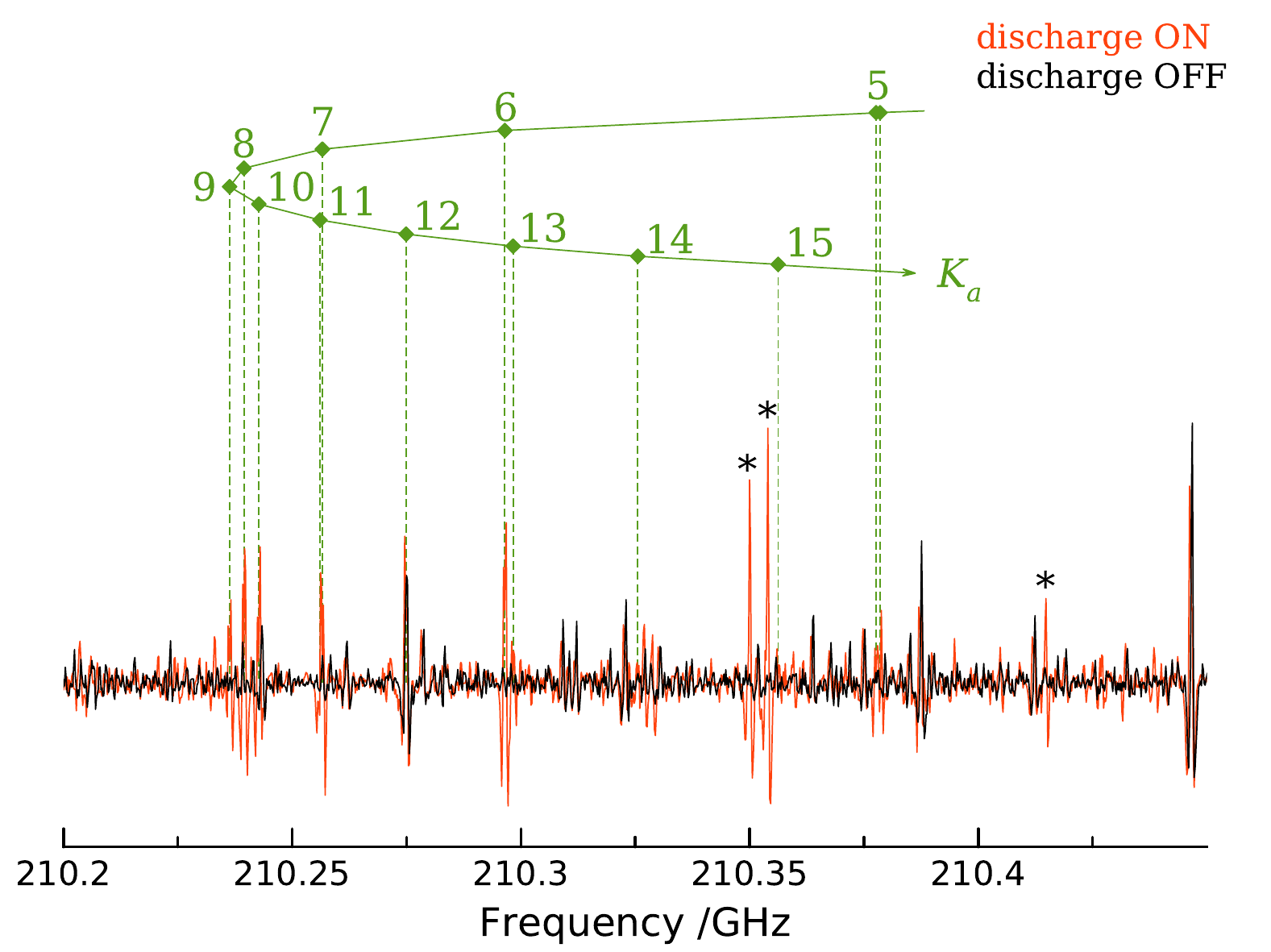}
\caption{Portion of the millimeter-wave spectrum recorded with RF discharge on (in red) and off (in black) showing the $K_a$ series in the $^qR(18)$ ($J' \leftarrow J'' = 19 \leftarrow 18$) branch of \textit{syn} vinyl mercaptan. For the lowest $K_a$ value visible on the plot, the asymmetry splitting ($K_a+K_c = J$ or $J+1$) is resolved. Transitions in black most likely arise from the precursor while transitions in red labeled with an asterisk remain unidentified (see text).}
\label{fig: broadSpectrum}
\end{figure}

\begin{figure}[ht!]
\centering
\includegraphics[width=0.49\textwidth]{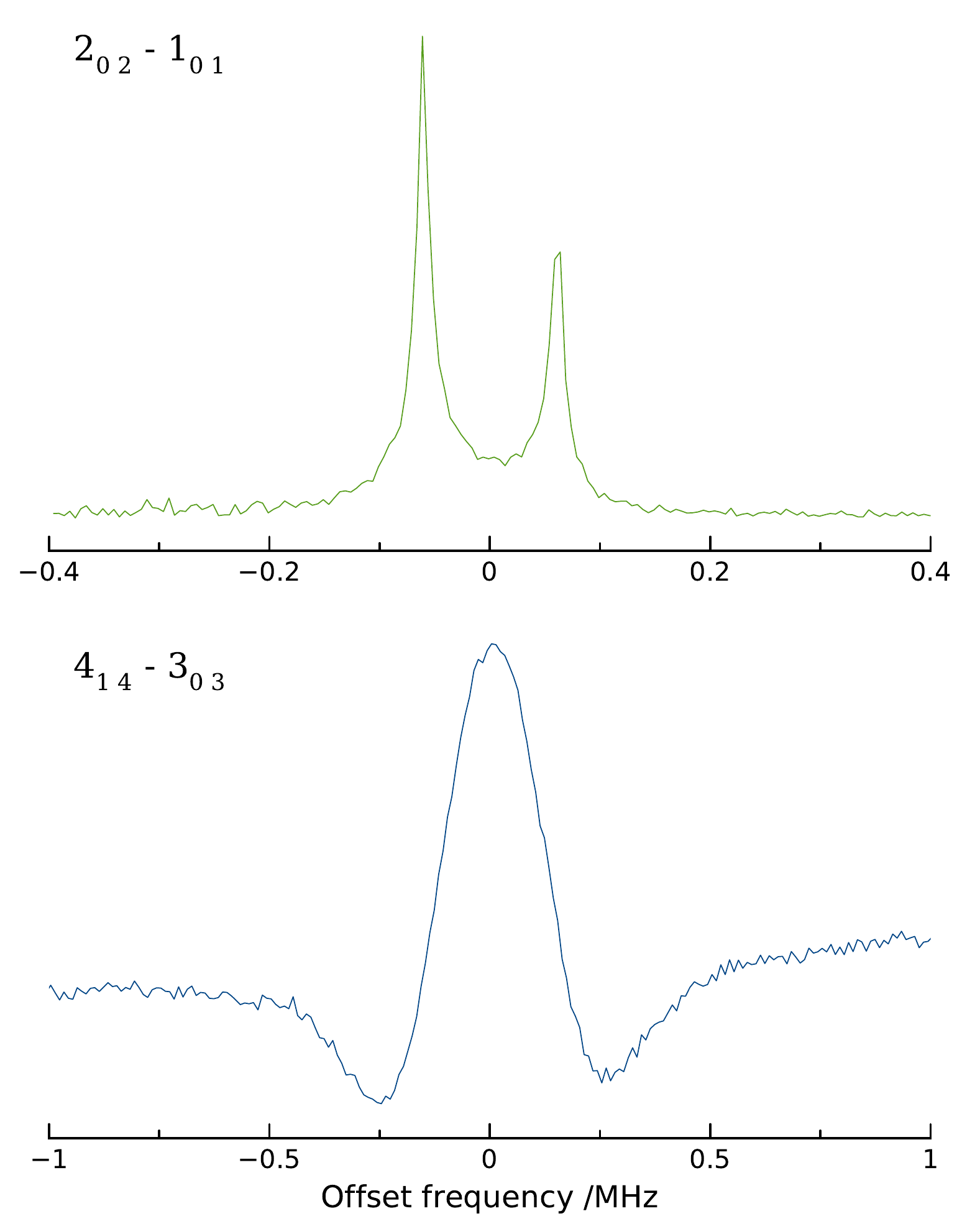}
\caption{Two transitions of \textit{anti} vinyl mercaptan recorded in the microwave (top trace, in green) and millimeter wave (bottom trace, in blue) domains. The frequency axis is given as an offset from the rest frequencies, 22346.6731 and 85467.985\,MHz, respectively. The doublet structure of the microwave transition (\textit{top spectrum}) is caused by Doppler splitting which arises because of the coaxial arrangement of the cavity axis with respect to the axis along which the gas expands; the rest frequency is taken as the average of the two components. The second derivative lineshape aspect of the millimeter transition (\textit{bottom spectrum}) arises from $2f$ demodulation of the modulated frequency.}
\label{fig: indivTransitions}
\end{figure}

Upon ignition of the discharge, relatively strong transitions of both \textit{syn} and \textit{anti} isomers of vinyl mercaptan were observed, as illustrated in Fig.~\ref{fig: broadSpectrum}. To derive rest frequencies as accurate as possible, rotational transitions have been measured individually using frequency steps of 10 and 20\,kHz in the 70 -- 120\,GHz and 170 -- 250\,GHz regions, respectively, and signal averaging was used to reach a signal-to-noise ratio of at least 10, and up to 50 in some cases (Fig. \ref{fig: indivTransitions}). The accuracy on each line frequency has been derived from a fit to the line profile, accounting for the full width at half maximum, the frequency step, and the signal-to-noise ratio as described in \cite{Martin-Drumel2015}.  Using this approach, rest frequencies  were derived with uncertainties as low as 10\,kHz.
Ultimately, nearly 300 and 150 new transitions (both $a$- and $b$-type) of \textit{syn} and \textit{anti} isomers of vinyl mercaptan have been recorded in the millimeter domain with quantum numbers up to $J''= 22$, $K_a'' = 17$ and  $J''= 19$, $K_a''= 14$, respectively. We note that, besides vinyl mercaptan, other species were produced under our experimental conditions (as indicated in Fig.\ref{fig: broadSpectrum}). Although some of their transitions were observed with high signal-to-noise ratios, they could not be assigned to any known species using the \rev{Cologne Database for Molecular Spectroscopy (CDMS)} and \rev{NASA Jet Propulsion Laboratory (JPL)} catalogs \citep{endres_cdms_2016, pickett_jpl}. The carriers of these lines may be vibrational satellites of vinyl mercaptan or its precursor, 1,2-ethanedithiol, or other species produced in the discharge.  Since no wide spectral surveys were conducted, however, no further line assignment attempts were made at this juncture.

\subsection{Centimeter wave spectroscopy}
For each isomer of vinyl mercaptan, roughly 10 transitions previously observed by \cite{Tanimoto1979} and \cite{Tanimoto1979a} have been re-measured  between 10 and 35\,GHz using the Cambridge Fourier-transform microwave spectrometer which has been described elsewhere \citep{mccarthy2000}. In doing so,  uncertainties on the rest frequencies of these lines have been reduced by nearly two orders of magnitude, from 100\,kHz to 1\,kHz. In this experiment, vinyl mercaptan was generated by a 1\,kV DC discharge in the throat of a small nozzle through a mixture of \ce{C2H2} and \ce{H2S} heavily diluted in neon and the resulting products were supersonically expanded into a large vacuum chamber.  By the time the products reach the center of the Fabry-Perot cavity,  where they are probed by the microwave radiation, the rotational temperature has fallen to 10\,K or less.  An example of a microwave transition of \textit{syn}-\ce{CH2CHSH} is presented in Fig. \ref{fig: indivTransitions}.

\subsection{Astronomical observations}
We searched for vinyl mercaptan in the EMoCA spectral line survey performed toward the Sgr B2(N) star-forming region with ALMA. We focus our analysis on the hot molecular core Sgr B2(N2) at the position ($\alpha, \delta_{{\rm J}2000}$) $=$ ($17^{\rm h}47^{\rm m}19.86^{\rm s}$, $-28^\circ22\arcmin13.4''$). Details about the 
observations, data reduction, and method used to identify the detected
lines and derive column densities can be found in \citet{Belloche16}. In 
short, the EMoCA survey covers the frequency range from 84.1\,GHz to 114.4\,GHz 
with a spectral resolution of 488.3\,kHz (1.7 to 1.3\,km\,s$^{-1}$). It was 
performed with five frequency tunings called S1 to S5. The median angular 
resolution is 1.6$''$.

\section{Results and discussion}

\subsection{Rest frequencies and derived spectroscopic parameters}
The transitions recorded in this study are summarized in Appendix \ref{ap:transitions}. 
From these measurements, and previously published data \citep{Tanimoto1979, Tanimoto1979a}, a fit to a Watson-$S$ Hamiltonian in the I$^r$ representation was performed using the SPFIT/SPCAT suite of programs \citep{Pickett1991}. The resulting best-fit spectroscopic constants are given in Table \ref{tab: param}. 

The large number of transitions recorded in this work coupled with the detection of lines that originate from much higher values of $J$ and $K_a$ compared to the previous studies enable centrifugal distortion (CD) parameters up to sixth-order to be precisely determined.  In some cases, the precision of these CD parameters has been improved by as much as four orders of magnitude. To reproduce the measured transitions to their experimental accuracy, a total of nine parameters are required, including two quartic CD terms ($D_K$ and $d_2$) and one sextic term ($H_{KJ}$) that were not determined previously.  The much larger dataset used in the present investigation and the  relatively small number of additional parameters needed point to the rapid convergence of the Hamiltonian model and the robustness of the fit. On the basis of the best-fit parameters in Table~\ref{tab: param}, the astronomically most interesting lines either have been measured or can now be predicted throughout the entire millimeter domain to better than 50 kHz up to 250\,GHz (0.06\,km/s in terms of \rev{radial} velocity).

\begin{table*}[ht!]
\caption{Ground state molecular parameters (in MHz) of \textit{syn} and \textit{anti} vinyl mercaptan derived in this study compared with previously published values \citep{Tanimoto1979, Tanimoto1979a} and those calculated theoretically \rev{(Calc.)}.}
\label{tab: param}
\newcolumntype{.}{D{.}{.}{-1}}
\centering

\footnotesize
\begin{tabular}{l...|...}
\toprule
 & \multicolumn{3}{c}{\textbf{\textit{syn}-\ce{CH2CHSH}}} & \multicolumn{2}{c}{\textbf{\textit{anti}-\ce{CH2CHSH}}} \\
 & \multicolumn{1}{c}{This work\tablefootmark{a}} & \multicolumn{1}{c}{Previous\tablefootmark{a}} & \multicolumn{1}{c}{Calc.\tablefootmark{b}} & \multicolumn{1}{c}{This work\tablefootmark{a}} & \multicolumn{1}{c}{Previous\tablefootmark{a}} & \multicolumn{1}{c}{Calc.\tablefootmark{b}}\\
\midrule
$A_e$                          & \multicolumn{1}{c}{$\cdots$} & \multicolumn{1}{c}{$\cdots$} & 50304.4 & \multicolumn{1}{c}{$\cdots$} & \multicolumn{1}{c}{$\cdots$} & 49843.5\\
$\Delta A_0$                   & \multicolumn{1}{c}{$\cdots$} & \multicolumn{1}{c}{$\cdots$} &   431.2 & \multicolumn{1}{c}{$\cdots$} & \multicolumn{1}{c}{$\cdots$} &    36.4\\
$B_e$                          & \multicolumn{1}{c}{$\cdots$} & \multicolumn{1}{c}{$\cdots$} &  5866.0 & \multicolumn{1}{c}{$\cdots$} & \multicolumn{1}{c}{$\cdots$} & 5912.5 \\
$\Delta B_0$                   & \multicolumn{1}{c}{$\cdots$} & \multicolumn{1}{c}{$\cdots$} &    38.1 & \multicolumn{1}{c}{$\cdots$} & \multicolumn{1}{c}{$\cdots$} &   31.9\\
$C_e$                          & \multicolumn{1}{c}{$\cdots$} & \multicolumn{1}{c}{$\cdots$} &  5253.4 & \multicolumn{1}{c}{$\cdots$} & \multicolumn{1}{c}{$\cdots$} & 5308.9 \\
$\Delta C_0$                   & \multicolumn{1}{c}{$\cdots$} & \multicolumn{1}{c}{$\cdots$} &    37.3 & \multicolumn{1}{c}{$\cdots$} & \multicolumn{1}{c}{$\cdots$} &   50.6 \\ \midrule
$A_0$                          &  49816.0376~(11) & 49815.28~(6)  & 49873.2 &  49423.5726~(11)  & 49422.75~(5)   &  49807.1 \\

$B_0$                          & 5835.708685~(92) & 5835.716~(14) &  5827.9 &  5897.21162~(11) & 5897.215~(9)   &   5880.6 \\ 

$C_0                         $ & 5222.075531~(85) & 5222.081~(11) &  5216.1 &  5279.43951~(13) & 5279.436~(9)   &   5258.3 \\ 
$D_{J}          \times 10^{3}$ &    2.724203~(89) & 2.85~(17)     &   2.716 &     3.09729~(17)  & 3.12~(11)      &      3.117   \\ 
$D_{JK}         \times 10^{3}$ &   -33.4874~(12) & -33.2~(21)    &  -35.89 &    -37.6145~(40)  & -38.5~(17)     &    -40.02    \\
$D_{K}                       $ &     0.79113~(14) & \multicolumn{1}{c}{$\cdots$} &  0.8100 &     0.80990~(18)  & \multicolumn{1}{c}{$\cdots$} &      0.9496  \\ 
$d_{1}          \times 10^{3}$ &   -0.424456~(88) & -0.425~(35)    & -0.4239 &    -0.47324~(19)  & -0.498~(51)~~~  &     -0.4143  \\ 
$d_{2}          \times 10^{3}$ &   -0.023739~(73) & \multicolumn{1}{c}{$\cdots$} & -0.0211 &    -0.03110~(25)  & \multicolumn{1}{c}{$\cdots$} &     -0.0411  \\  
$H_{KJ}         \times 10^{6}$ &     -3.1367~(59) & \multicolumn{1}{c}{$\cdots$} & \multicolumn{1}{c|}{$\cdots$} &      -4.356~(31)  & \multicolumn{1}{c}{$\cdots$} & \multicolumn{1}{c}{$\cdots$} \\
\midrule
\# of lines  & \multicolumn{1}{c}{333} &  \multicolumn{1}{c}{37}&  \multicolumn{1}{c|}{$\cdots$} & \multicolumn{1}{c}{205}  &\multicolumn{1}{c}{32} & \multicolumn{1}{c}{$\cdots$} \\ 
RMS /kHz  &  \multicolumn{1}{c}{43} &  \multicolumn{1}{c}{$\cdots$}  & \multicolumn{1}{c|}{$\cdots$}  &  \multicolumn{1}{c}{49}  & \multicolumn{1}{c}{$\cdots$}  &  \multicolumn{1}{c}{$\cdots$} \\
$\sigma$\tablefootmark{c}  & \multicolumn{1}{c}{0.95}&  \multicolumn{1}{c}{$\cdots$}  & \multicolumn{1}{c|}{$\cdots$}  &\multicolumn{1}{c}{1.03} & \multicolumn{1}{c}{$\cdots$}   &  \multicolumn{1}{c}{$\cdots$} \\ 
\bottomrule
\end{tabular}
\tablefoot{
\tablefoottext{a}{Numbers in parentheses are 1$\sigma$ uncertainties in units of the last digit.}
\tablefoottext{b}{$A_e$, $B_e$, and $C_e$ rotational constants derived from structure calculated at the  ae-CCSD(T)/cc-pwCVQZ level of theory. Zero-point vibrational corrections $\Delta A_0$, $\Delta B_0$, and $\Delta C_0$ and quartic centrifugal distortion constants were calculated at the fc-CCSD(T)/cc-pV(T+$d$)Z level.}
\tablefoottext{c}{Reduced standard deviation (unitless).}
}
\end{table*}

\subsection{On the poor predictive power of extrapolating high-frequency transitions from microwave measurements}

\begin{figure}[ht!]
\centering
\includegraphics[width=0.49\textwidth]{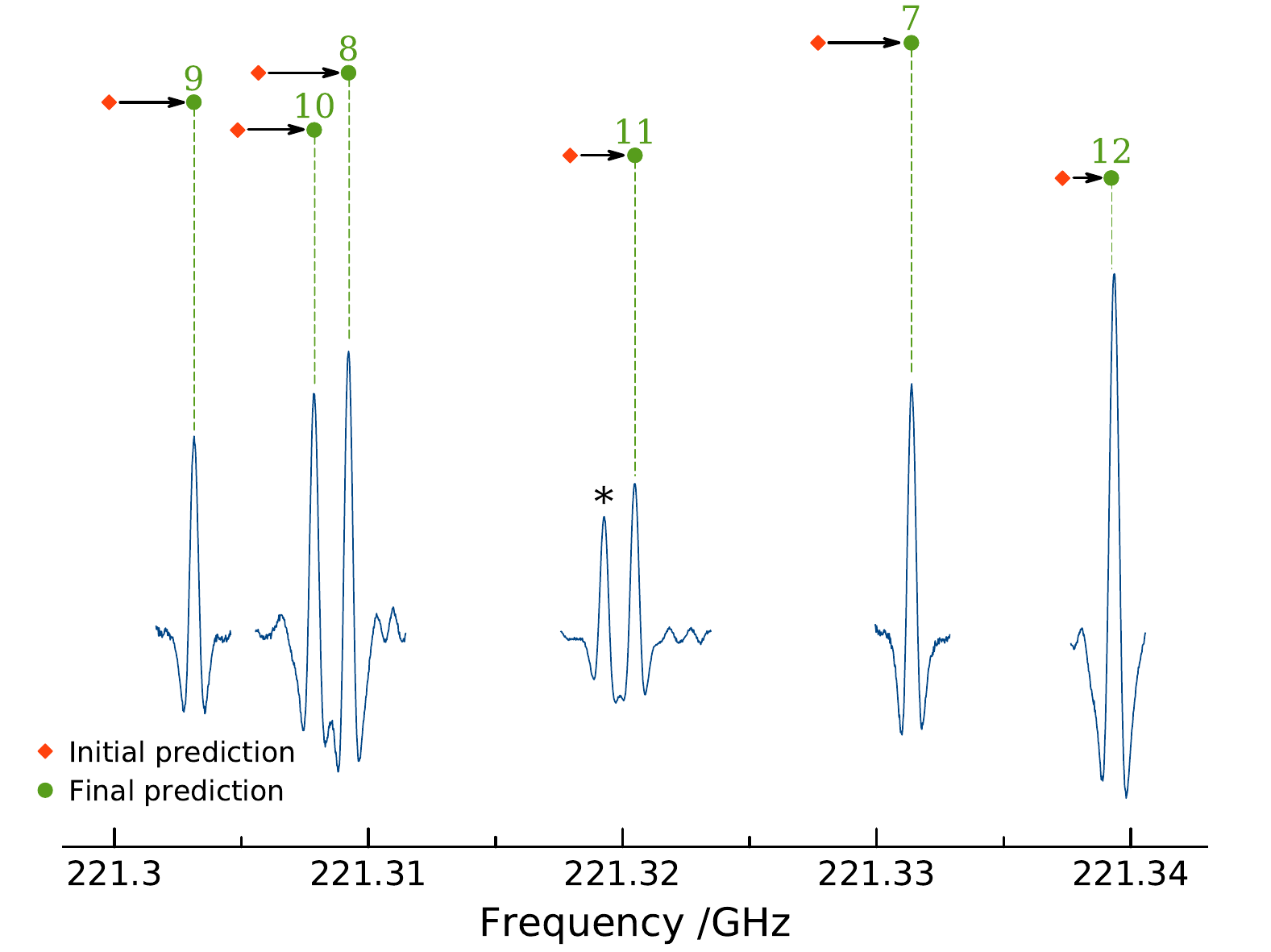}
\caption{Individual transitions measured in the $^qR(19)$ branch of \textit{syn} vinyl mercaptan and comparison between the initial (red diamonds) and final (green dots) predictions. The $K_a$ value of each transition is indicated in green; the asymmetry splitting is not resolved for these lines. The transition labeled with an asterisk is unidentified. The lineshape appears as a second derivative owing to the employed detection scheme.}
\label{fig: deviation}
\end{figure}

While frequency predictions based on the spectroscopic parameters from \cite{Tanimoto1979} and \cite{Tanimoto1979a} have enabled many new lines to be assigned  in the  millimeter-wave band, this parameter set alone is inadequate to detect lines of vinyl mercaptan in astronomical sources. Many high-$J$, high-$K_a$ transitions, for example, were shifted from their initial predictions by a few to up to tens of MHz (e.g., the transition $16_{3\, 14} - 16_{2\, 15}$  initially predicted at 226581.2\,MHz was measured at 226532.7\,MHz). In space, the most relevant transitions are the $^qR$ series for the \textit{syn} isomer ($\Delta J =1$, $\Delta K_a = 0$) as these are the strongest lines at temperatures ranging from a few tens to a few hundreds of K. For these lines, the typical deviation from the initial prediction was in the range of 2--5\,MHz around 200\,GHz (Fig. \ref{fig: deviation}), corresponding to a significant fraction of the 2--10\,km/s linewidths that typify hot cores and corinos in star forming regions.
This example illustrates the pitfalls of extrapolating from the microwave to the millimeter and submillimeter domain, particularly for non-rigid molecules, and emphasizes the value of  high-frequency laboratory measurements of astrophysically relevant species, even when microwave spectrum data is already available.

\subsection{Isomers}

The relative stability of isomers may be an important factor in predicting  isomeric abundances in the ISM.   
With this motivation in mind, we performed high-level \textit{ab initio} calculations (Fig. \ref{fig:relativeenergies}) to determine the relative stability of [\ce{C2},\ce{H4},S] isomers and the corresponding isovalent oxygen analogs; for the former, the calculations presented here are --- to the knowledge of the authors --- the first report of the relative energetics at 0\,K. Because our calculations for
the [\ce{C2},\ce{H4},O] isomeric family are in excellent agreement with experiments, we can in turn  be confident of the [\ce{C2},\ce{H4},S] thermochemistry which is significantly less well constrained by measurements. 

In  contrast to the oxygen isomers, the four lowest single [\ce{C2},\ce{H4},S] isomers are effectively isoenergetic at our current level of theoretical sophistication. While it is unclear if a higher level of treatment (requiring sub-kJ/mol) will change the energetic ordering of these isomers, the near-degeneracy of [\ce{C2},\ce{H4},S] isomers presents a fairly unique test bed to assess possible reaction pathways in astrochemical objects. The role of icy grains in [\ce{C2},\ce{H4},O] formation has been suggested as the underlying factor as to why abundance ratios depart from thermodynamic equilibrium \citep{bennett_laboratory_2005}. Having established that both isomers of vinyl mercaptan and thiirane are effectively isoenergetic, their relative interstellar abundance should present a more sensitive probe of local processes (e.g., shock and grain chemistry) than their oxygen analogs, which are more widely spaced in energy (Fig. \ref{fig:relativeenergies}). A systematic astronomical investigation of the [\ce{C2},\ce{H4},S] family of isomers, however, is currently limited by the lack of reliable frequencies in the millimeter and submillimeter domains for thioacetaldehyde, the remaining isomer for which no data has presently been published beyond the microwave domain.

\subsection{Astronomical searches}
Searches for the \textit{syn} and \textit{anti} isomers of vinyl mercaptan were performed with the EMoCA spectrum of Sgr B2(N2) using the best-fit parameters in Table \ref{tab: param}. Weeds \citep[][]{Maret11} was used to produce  synthetic spectra of both isomers under the \rev{local thermodynamic equilibrium (LTE)} approximation, and the same parameters (emission size, rotational temperature, linewidth, velocity offset with respect to the assumed systemic velocity of the source) as the ones derived for methyl mercaptan in \citet{Mueller16} were assumed here. Hence, the only free parameters are the column densities of the \textit{syn} and \textit{anti} forms of vinyl mercaptan,
which we consider as independent species. 

No evidence was found for either isomer in the EMoCA spectrum. The upper limits of the  column densities are reported in Table~\ref{t:coldens}, along with the column densities or upper limits thereof derived for other molecules in our previous EMoCA studies \citep[][]{Mueller16,Belloche16}. In addition, no evidence was found for the \textit{syn} and \textit{anti} isomers of vinyl alcohol  using predictions derived from the pure rotational works reported in \cite{Saito1976} and \cite{RODLER1985}. The corresponding upper limits of their column densities are summarized  in Table~\ref{t:coldens} as well. The non-detection of vinyl alcohol with ALMA along with its non-detection with the IRAM 30\,m telescope \citep{Belloche2013} confirms that if vinyl alcohol were present in the Sgr B2(N) star forming region at the level reported by \cite{Turner2001}\rev{, whose observations with the Kitt Peak 12 m telescope did not resolve the multiple hot cores embedded in Sgr B2(N)}, then its emission must arise from the source envelope on large scales ($> 1$\,pc) and not from the much more compact hot cores, as already stated by \cite{Belloche2013}. Surveys like PRIMOS \rev{(Prebiotic Interstellar Molecular Survey, \cite{Neill2012})}  or the one performed at the \rev{Arizona Radio Observatory (ARO)} \citep{Halfen2017} may shed more light on the presence and distribution of vinyl alcohol  toward Sgr B2(N).

\begin{table*}[ht!]
 \caption{ Parameters of our best-fit LTE model (or upper limit) of selected complex organic molecules toward Sgr~B2(N2). }
\label{t:coldens}
 \vspace*{-2.5ex}
 \begin{center}
 \begin{tabular}{lcrccccccrc}
 \hline\hline
 \multicolumn{1}{c}{Molecule} & \multicolumn{1}{c}{Status\tablefootmark{a}} & \multicolumn{1}{c}{$N_{\rm det}$\tablefootmark{b}} & \multicolumn{1}{c}{Size\tablefootmark{c}} & \multicolumn{1}{c}{$T_{\mathrm{rot}}$\tablefootmark{d}} & \multicolumn{1}{c}{$N$\tablefootmark{e}} & \multicolumn{1}{c}{$C$\tablefootmark{f}} & \multicolumn{1}{c}{$\Delta V$\tablefootmark{g}} & \multicolumn{1}{c}{$V_{\mathrm{off}}$\tablefootmark{h}} & \multicolumn{1}{c}{$\frac{N_{\rm ref}}{N}$\tablefootmark{i}} & \multicolumn{1}{c}{Ref.\tablefootmark{j}} \\ 
  & & & \multicolumn{1}{c}{\scriptsize ($''$)} & \multicolumn{1}{c}{\scriptsize (K)} & \multicolumn{1}{c}{\scriptsize (cm$^{-2}$)} & & \multicolumn{1}{c}{\scriptsize (km~s$^{-1}$)} & \multicolumn{1}{c}{\scriptsize (km~s$^{-1}$)} & & \\ 
 \hline
 CH$_3$SH, $\varv=0^\star$ & d & 12 & 1.40 &  180 &  3.4 (17) & 1.00 & 5.4 & $-0.5$ &       1 & 2 \\ 
 \textit{gauche}-C$_2$H$_5$SH & n & 0 & 1.40 &  180 & $<$  1.6 (16) & 1.95 & 5.4 & $-0.5$ & $>$      22 & 2 \\ 
 \textit{syn}-C$_2$H$_3$SH & n & 0 & 1.40 &  180 & $<$  3.7 (16) & 1.22 & 5.4 & $-0.5$ & $>$       9 & 1 \\ 
 \textit{anti}-C$_2$H$_3$SH & n & 0 & 1.40 &  180 & $<$  1.6 (16) & 1.63 & 5.4 & $-0.5$ & $>$      21 & 1 \\ 
\hline 
 CH$_3$CN, $\varv_8=1^\star$ & d & 20 & 1.40 &  170 &  2.2 (18) & 1.00 & 5.4 & $-0.5$ &       1 & 3 \\ 
 C$_2$H$_5$CN, $\varv=0$ & d & 154 & 1.20 &  150 &  6.2 (18) & 1.38 & 5.0 & $-0.8$ &    0.35 & 3 \\ 
 C$_2$H$_3$CN, $\varv=0$ & d & 44 & 1.10 &  200 &  4.2 (17) & 1.00 & 6.0 & $-0.6$ &       5 & 3 \\ 
\hline 
 CH$_3$OH, $\varv_{\rm t}=1^\star$ & d & 16 & 1.40 &  160 &  4.0 (19) & 1.00 & 5.4 & $-0.2$ &       1 & 2 \\ 
 C$_2$H$_5$OH & d & 168 & 1.25 &  150 &  2.0 (18) & 1.24 & 5.4 & $0.0$ &      20 & 2 \\ 
 \textit{syn}-C$_2$H$_3$OH & n & 0 & 1.25 &  150 & $<$  4.0 (16) & 1.00 & 5.4 & $0.0$ & $>$    1000 & 1 \\ 
 \textit{anti}-C$_2$H$_3$OH & n & 0 & 1.25 &  150 & $<$  1.2 (16) & 1.00 & 5.4 & $0.0$ & $>$    3333 & 1 \\ 
\hline 
 \end{tabular}
 \end{center}
 \vspace*{-2.5ex}
 \tablefoot{
 \tablefoottext{a}{d: detection, n: non-detection.}
 \tablefoottext{b}{Number of detected lines \citep[conservative estimate, see Sect.~3 of][]{Belloche16}. One line of a given species may mean a group of transitions of that species that are blended together.}
 \tablefoottext{c}{Emission diameter (\rev{full width at half maximum,}  \textit{FWHM}).}
 \tablefoottext{d}{Rotational temperature.}
 \tablefoottext{e}{Total column density of the molecule, except for the \textit{syn} and \textit{trans} conformers of vinyl mercaptan and vinyl alcohol which are in each case considered as independent species. For these species, the column density represents the column density of each conformer separately. $X$ ($Y$) means $X \times 10^Y$.}
 \tablefoottext{f}{Correction factor that was applied to the column density to account for the contribution of vibrationally or torsionally excited states or other conformers, in the cases where this contribution was not included in the partition function of the spectroscopic predictions.}
 \tablefoottext{g}{Linewidth (\textit{FWHM}).}
 \tablefoottext{h}{Velocity offset with respect to the assumed systemic velocity of Sgr~B2(N2), $V_{\mathrm{lsr}} = 74$ km~s$^{-1}$.}
 \tablefoottext{i}{Column density ratio, with $N_{\rm ref}$ the column density of the previous reference species marked with a $\star$.}
 \tablefoottext{j}{References: (1) this work, (2) \citet{Mueller16}, (3) \citet{Belloche16}.}
 }
 \end{table*}

We find that the \textit{syn} and \textit{anti} isomers of vinyl mercaptan are at least 9 and 21 times less abundant than methyl mercaptan, respectively. The energy difference between the \textit{syn} and \textit{anti} isomers expressed in temperature units is about 60\,K, which implies that the abundance ratio between the \textit{anti} and \textit{syn} isomers is 0.72 for a temperature of 180\,K. This means that the total column density of vinyl mercaptan at this temperature is 1.72 times that of the \textit{syn} isomer or equivalently 2.4 times that of the \textit{anti} isomer. The most stringent constraint on the total column density of vinyl mercaptan is provided by the \textit{anti} isomer: vinyl mercaptan is at least $\sim 9$ times (21/2.4) less abundant than methyl mercaptan.

The  lower limit for the abundance ratio of vinyl to methyl mercaptan can be compared to ratios obtained for other families of molecules. It is nearly a factor of two higher than the abundance ratio measured for the cyanides ([CH$_3$CN]/[C$_2$H$_3$CN] $=$ 5). For vinyl alcohol, the large energy difference
between the \textit{syn} and \textit{anti} isomers (4.6\,kJ/mol, i.e.  385\,cm$^{-1}$, Fig. \ref{fig:relativeenergies}) implies that the total column density of vinyl alcohol is equal to 1.025 times that of the \textit{syn} isomer at 150\,K. We thus conclude that vinyl alcohol is at least 1000 times less abundant than methanol. This lower limit is more than two orders of magnitude higher than the ratio measured for the cyanides. Given that the abundance ratio of methanol to ethanol is also higher (by a factor $\sim 60$) than the abundance ratio of the corresponding cyanides, the chemistries of the cyanides and alcohols appear to differ significantly. However, the observational constraints do not tell us whether the chemistries of the mercaptans and alcohols are similar or differ: the lower limit to the ratio of methyl to ethyl mercaptan is marginally consistent with the corresponding ratio measured for the alcohols, and the lower limits to the ratio of the methyl to vinyl species do not exclude a similar ratio in both families.

\section{Conclusion}

In the present study, a rotational investigation of the two isomers of vinyl mercaptan has been extended up to 250\,GHz. The new set of transition frequencies and derived spectroscopic constants provide a highly reliable basis for subsequent searches for both species in the ISM. Calculations of the rotational spectra are available in the catalog section of the CDMS \citep{endres_cdms_2016} database\footnote{\texttt{https://www.astro.uni-koeln.de/cdms/catalog}} while linelists, parameters, fits and auxiliary files will be available in the data section\footnote{\texttt{https://www.astro.uni-koeln.de/cdms/daten}}.

The work performed in this study highlights the urgent need for high frequency measurements of numerous species of interstellar interest that so far have only been characterized at centimeter wavelengths: indeed, most of the current ISM observations are now performed in the millimeter and submillimeter domains, using facilities such as ALMA, NOEMA, APEX, and SOFIA, but frequency extrapolations into the millimeter domain using spectroscopic constants derived from low $J$ and $K_a$ observations are inherently uncertain, and make detection of new astronomical species unduly challenging. Additionally, this work provides \rev{constraints} on isomeric populations in the ISM. Both \textit{syn} and \textit{anti} vinyl mercaptan have been the subjects of searches in the EMoCA survey recorded toward the hot core Sgr B2(N2). No transitions of either species has been detected and upper limits for their column densities have been derived, revealing that vinyl mercaptan is at least 9 times less abundant than methyl mercaptan in this source.

\begin{acknowledgements}
      Part of this work was supported by the German \emph{Deutsche Forschungsgemeinschaft} (DFG)  collaborative research grant SFB~956 (project B3), SCHL 341/15-1 (Cologne Center for Terahertz Spectroscopy), and the NSF grant AST-1615847. M.-A. M.-D. is thankful to the Programme National ``Physique et Chimie du Milieu Interstellaire'' (PCMI) of CNRS/INSU with INC/INP co-funded by CEA and CNES for support.  M. C. M. thanks NASA grants NNX13AE59G and 80NSSC18K0396.
\end{acknowledgements}

\clearpage
%
\bibliographystyle{aa} 
\bibliography{Bib} 
%

\clearpage
\appendix

\section{Molecular structures of the [\ce{C2},\ce{H4},S] isomers \label{ap:zmat}}

\small
\begin{verbatim}
Internal coordinates ae-CCSD(T)/cc-pwCVQZ

syn vinyl mercaptan 
H
C 1 r1
C 2 r2 1 a1
S 3 r3 2 a2 1 d0
H 4 r4 3 a3 2 d0
H 2 r5 3 a4 1 d180
H 3 r6 2 a5 6 d0

r1   =        1.08115
r2   =        1.33270
a1   =      122.05669
r3   =        1.75221
a2   =      127.33692
d0   =        0.00000
r4   =        1.33663
a3   =       96.56932
r5   =        1.07962
a4   =      120.07569
d180 =      180.00000
r6   =        1.08206
a5   =      120.75551

anti vinyl mercaptan 
H
C 1 r1
C 2 r2 1 a1
S 3 r3 2 a2 1 d1
H 4 r4 3 a3 2 d2
H 2 r5 3 a4 1 d3
H 3 r6 2 a5 6 d4

r1   =        1.08105
r2   =        1.33250
a1   =      121.79666
r3   =        1.75849
a2   =      122.52864
d1   =        3.65449
r4   =        1.33431
a3   =       96.59150
d2   =     -159.42502
r5   =        1.07959
a4   =      120.21684
d3   =     -179.52134
r6   =        1.08069
a5   =      121.24236
d4   =        0.62115

thiirane
S
C 1 r1
C 1 r1 1 a1
H 2 r2 1 a2 3 d1
H 2 r2 1 a2 3 md1
H 3 r2 1 a2 2 d1
H 3 r2 1 a2 2 md1

r1   =        1.81067
a1   =       48.27245
r2   =        1.07991
a2   =      115.09725
d1   =      110.93875
md1  =     -110.93875

thioacetaldehyde
S
C 1 r1
C 2 r2 1 a1
H 3 r3 2 a2 1 d0
H 2 r4 3 a3 1 d180
H 3 r5 2 a4 5 d1
H 3 r5 2 a4 5 md1

r1   =        1.61419
r2   =        1.49172
a1   =      125.71288
r3   =        1.08597
a2   =      111.24050
d0   =        0.00000
r4   =        1.08887
a3   =      115.25385
d180 =      180.00000
r5   =        1.09181
a4   =      109.69777
d1   =       58.47364
md1  =      -58.47364
\end{verbatim}
\newpage

\section{Observed experimental transitions of vinyl mercaptan\label{ap:transitions}}

\vspace{1cm}

\newcolumntype{.}{D{.}{.}{-1}}
\tablehead{
\toprule
 $J'$ & $K'_a$ & $K'_c$ & $J''$ & $K''_a$ & $K''_c$ & \multicolumn{1}{c}{Obs.} & \multicolumn{1}{c}{Obs - Calc.} \\
\midrule}
\tabletail{\bottomrule}
\topcaption{Observed frequencies for \textit{syn}-vinyl mercaptan. Frequencies are given in MHz. Values in parentheses represent $1\sigma$ uncertainty. The obs. -- calc. values are given in kHz.}
\begin{supertabular}{c c c c c c . .}
1 & 0 & 1 & 0 & 0 & 0 & 11,057.7730(10) & -0.3\\
2 & 1 & 2 & 1 & 1 & 1 & 21,501.9964(10) & 0.8\\
1 & 1 & 1 & 2 & 0 & 2 & 21,870.5017(10) & -0.2\\
2 & 0 & 2 & 1 & 0 & 1 & 22,109.1044(10) & -0.1\\
2 & 1 & 1 & 1 & 1 & 0 & 22,729.2350(10) & 0.2\\
3 & 1 & 3 & 2 & 1 & 2 & 32,248.8905(10) & 0.3\\
3 & 0 & 3 & 2 & 0 & 2 & 33,147.5585(10) & 1.3\\
3 & 1 & 2 & 2 & 1 & 1 & 34,089.6590(20) & 2.4\\
7 & 3 & 5 & 6 & 3 & 4 & 77,455.364(15) & 0.5\\
7 & 3 & 4 & 6 & 3 & 3 & 77,460.551(20) & -21.0\\
7 & 2 & 5 & 6 & 2 & 4 & 77,707.694(10) & 1.9\\
7 & 1 & 6 & 6 & 1 & 5 & 79,462.064(10) & 8.6\\
8 & 1 & 8 & 7 & 1 & 7 & 85,881.234(10) & 4.6\\
8 & 0 & 8 & 7 & 0 & 7 & 87,927.042(10) & 0.7\\
8 & 2 & 7 & 7 & 2 & 6 & 88,380.994(10) & -0.5\\
8 & 6 & 2 & 7 & 6 & 1 & 88,499.974(10) & 4.6\\
8 & 6 & 3 & 7 & 6 & 2 & 88,499.974(10) & 4.6\\
8 & 5 & 3 & 7 & 5 & 2 & 88,501.223(10) & 2.1\\
8 & 5 & 4 & 7 & 5 & 3 & 88,501.223(10) & 2.4\\
8 & 4 & 4 & 7 & 4 & 3 & 88,510.009(10) & -36.2\\
8 & 4 & 5 & 7 & 4 & 4 & 88,510.009(10) & 37.3\\
8 & 3 & 6 & 7 & 3 & 5 & 88,532.822(10) & -10.2\\
8 & 3 & 5 & 7 & 3 & 4 & 88,543.245(10) & 4.9\\
8 & 2 & 6 & 7 & 2 & 5 & 88,910.481(10) & 2.4\\
8 & 1 & 7 & 7 & 1 & 6 & 90,778.139(10) & -8.9\\
8 & 1 & 7 & 7 & 1 & 6 & 90,778.139(10) & -8.9\\
15 & 1 & 14 & 15 & 0 & 15 & 91,165.379(10) & -6.1\\
9 & 1 & 9 & 8 & 1 & 8 & 96,577.987(15) & 13.3\\
16 & 1 & 15 & 16 & 0 & 16 & 98,557.507(20) & 20.6\\
9 & 0 & 9 & 8 & 0 & 8 & 98,760.335(10) & -5.9\\
9 & 2 & 8 & 8 & 2 & 7 & 99,399.930(10) & 1.2\\
9 & 6 & 3 & 8 & 6 & 2 & 99,565.434(10) & 3.5\\
9 & 6 & 4 & 8 & 6 & 3 & 99,565.434(10) & 3.5\\
9 & 5 & 5 & 8 & 5 & 4 & 99,568.972(10) & 6.6\\
9 & 5 & 4 & 8 & 5 & 3 & 99,568.972(10) & 5.8\\
9 & 4 & 6 & 8 & 4 & 5 & 99,582.945(15) & 115.8\\
9 & 4 & 5 & 8 & 4 & 4 & 99,582.945(15) & -60.4\\
9 & 3 & 7 & 8 & 3 & 6 & 99,614.458(10) & 2.8\\
9 & 3 & 6 & 8 & 3 & 5 & 99,633.511(10) & -2.6\\
9 & 1 & 8 & 8 & 1 & 7 & 102,079.135(10) & -2.5\\
12 & 0 & 12 & 11 & 1 & 11 & 103,466.696(10) & -9.9\\
17 & 1 & 16 & 17 & 0 & 17 & 106,576.954(10) & 0.9\\
10 & 0 & 10 & 9 & 0 & 9 & 109,542.337(10) & 2.6\\
10 & 2 & 9 & 9 & 2 & 8 & 110,408.779(10) & 1.2\\
10 & 6 & 4 & 9 & 6 & 3 & 110,631.942(10) & 5.6\\
10 & 6 & 5 & 9 & 6 & 4 & 110,631.942(10) & 5.6\\
10 & 5 & 6 & 9 & 5 & 5 & 110,638.505(10) & -0.2\\
10 & 5 & 5 & 9 & 5 & 4 & 110,638.505(10) & -2.4\\
10 & 4 & 7 & 9 & 4 & 6 & 110,658.876(10) & 7.9\\
10 & 4 & 7 & 9 & 4 & 6 & 110,658.876(10) & 7.9\\
10 & 4 & 7 & 9 & 4 & 6 & 110,658.876(10) & 7.9\\
10 & 4 & 6 & 9 & 4 & 5 & 110,659.239(10) & -10.7\\
10 & 3 & 8 & 9 & 3 & 7 & 110,700.256(10) & -5.3\\
10 & 3 & 7 & 9 & 3 & 6 & 110,732.885(10) & 4.5\\
10 & 2 & 8 & 9 & 2 & 7 & 111,432.824(10) & 4.9\\
10 & 1 & 9 & 9 & 1 & 8 & 113,362.602(10) & 0.2\\
13 & 0 & 13 & 12 & 1 & 12 & 116,441.566(10) & -10.0\\
11 & 1 & 11 & 10 & 1 & 10 & 117,933.385(10) & -0.8\\
11 & 0 & 11 & 10 & 0 & 10 & 120,270.722(10) & -6.2\\
11 & 2 & 10 & 10 & 2 & 9 & 121,406.446(10) & -7.2\\
11 & 4 & 8 & 10 & 4 & 7 & 121,738.377(30) & -36.4\\
11 & 4 & 7 & 10 & 4 & 6 & 121,739.165(10) & -10.9\\
16 & 1 & 16 & 15 & 1 & 15 & 171,074.952(10) & -19.3\\
16 & 0 & 16 & 15 & 0 & 15 & 173,142.577(10) & 4.5\\
16 & 2 & 15 & 15 & 2 & 14 & 176,190.615(10) & -9.5\\
16 & 8 & 9 & 15 & 8 & 8 & 177,033.988(50) & 40.1\\
16 & 8 & 8 & 15 & 8 & 7 & 177,033.988(50) & 40.1\\
16 & 9 & 7 & 15 & 9 & 6 & 177,037.310(10) & 10.8\\
16 & 9 & 8 & 15 & 9 & 7 & 177,037.310(10) & 10.8\\
16 & 7 & 9 & 15 & 7 & 8 & 177,039.557(10) & -9.0\\
16 & 7 & 10 & 15 & 7 & 9 & 177,039.557(10) & -9.0\\
16 & 10 & 6 & 15 & 10 & 5 & 177,046.927(10) & -5.2\\
16 & 10 & 7 & 15 & 10 & 6 & 177,046.927(10) & -5.2\\
16 & 6 & 11 & 15 & 6 & 10 & 177,059.365(10) & 9.3\\
16 & 6 & 10 & 15 & 6 & 9 & 177,059.365(10) & 7.6\\
16 & 11 & 5 & 15 & 11 & 4 & 177,061.372(30) & 37.7\\
16 & 11 & 6 & 15 & 11 & 5 & 177,061.372(30) & 37.7\\
16 & 12 & 4 & 15 & 12 & 3 & 177,079.577(10) & -17.0\\
16 & 12 & 5 & 15 & 12 & 4 & 177,079.577(10) & -17.0\\
16 & 13 & 3 & 15 & 13 & 2 & 177,101.184(30) & 55.0\\
16 & 13 & 4 & 15 & 13 & 3 & 177,101.184(30) & 55.0\\
16 & 5 & 11 & 15 & 5 & 10 & 177,104.608(10) & -88.7\\
16 & 5 & 12 & 15 & 5 & 11 & 177,104.608(10) & 92.1\\
16 & 14 & 3 & 15 & 14 & 2 & 177,125.537(90) & -10.7\\
16 & 14 & 2 & 15 & 14 & 1 & 177,125.537(90) & -10.7\\
16 & 4 & 13 & 15 & 4 & 12 & 177,198.676(10) & -8.3\\
16 & 4 & 12 & 15 & 4 & 11 & 177,209.792(10) & 7.2\\
16 & 3 & 14 & 15 & 3 & 13 & 177,284.117(10) & 6.7\\
16 & 3 & 13 & 15 & 3 & 12 & 177,628.437(10) & -11.8\\
16 & 2 & 14 & 15 & 2 & 13 & 179,916.094(10) & 5.0\\
16 & 1 & 15 & 15 & 1 & 14 & 180,534.684(10) & 10.2\\
17 & 2 & 16 & 17 & 1 & 17 & 181,199.29(12) & -7.0\\
17 & 1 & 17 & 16 & 1 & 16 & 181,659.664(20) & -34.1\\
17 & 0 & 17 & 16 & 0 & 16 & 183,591.361(10) & -3.9\\
18 & 2 & 17 & 18 & 1 & 18 & 186,959.83(10) & -22.8\\
17 & 2 & 16 & 16 & 2 & 15 & 187,099.739(10) & -11.5\\
17 & 8 & 9 & 16 & 8 & 8 & 188,101.835(10) & -11.5\\
17 & 8 & 10 & 16 & 8 & 9 & 188,101.835(10) & -11.5\\
17 & 9 & 9 & 16 & 9 & 8 & 188,103.400(10) & -4.7\\
17 & 9 & 8 & 16 & 9 & 7 & 188,103.400(10) & -4.7\\
17 & 7 & 11 & 16 & 7 & 10 & 188,110.757(10) & -6.1\\
17 & 7 & 10 & 16 & 7 & 9 & 188,110.757(10) & -6.1\\
17 & 10 & 8 & 16 & 10 & 7 & 188,112.210(10) & -5.8\\
17 & 10 & 7 & 16 & 10 & 6 & 188,112.210(10) & -5.8\\
17 & 11 & 7 & 16 & 11 & 6 & 188,126.474(15) & 4.9\\
17 & 11 & 6 & 16 & 11 & 5 & 188,126.474(15) & 4.9\\
17 & 6 & 12 & 16 & 6 & 11 & 188,136.376(10) & -18.2\\
17 & 6 & 11 & 16 & 6 & 10 & 188,136.376(10) & -21.7\\
17 & 12 & 5 & 16 & 12 & 4 & 188,145.089(10) & 14.3\\
17 & 12 & 6 & 16 & 12 & 5 & 188,145.089(10) & 14.3\\
17 & 13 & 4 & 16 & 13 & 3 & 188,167.379(50) & 40.8\\
17 & 13 & 5 & 16 & 13 & 4 & 188,167.379(50) & 40.8\\
17 & 5 & 13 & 16 & 5 & 12 & 188,192.280(20) & 137.2\\
17 & 5 & 12 & 16 & 5 & 11 & 188,192.280(20) & -178.7\\
17 & 4 & 14 & 16 & 4 & 13 & 188,304.785(10) & -14.6\\
17 & 4 & 13 & 16 & 4 & 12 & 188,321.821(10) & -9.6\\
17 & 3 & 15 & 16 & 3 & 14 & 188,386.333(10) & 17.2\\
17 & 3 & 14 & 16 & 3 & 13 & 188,849.934(10) & -13.2\\
17 & 2 & 15 & 16 & 2 & 14 & 191,419.205(10) & -11.9\\
17 & 1 & 16 & 16 & 1 & 15 & 191,610.819(10) & -12.6\\
19 & 0 & 19 & 18 & 1 & 18 & 191,859.479(25) & 12.8\\
18 & 1 & 18 & 17 & 1 & 17 & 192,230.329(10) & -11.9\\
18 & 0 & 18 & 17 & 0 & 17 & 194,012.562(10) & -16.9\\
6 & 2 & 5 & 5 & 1 & 4 & 194,594.052(85) & -53.9\\
17 & 1 & 17 & 16 & 0 & 16 & 197,926.936(15) & -5.6\\
18 & 2 & 17 & 17 & 2 & 16 & 197,990.896(10) & -1.8\\
18 & 9 & 9 & 17 & 9 & 8 & 199,169.747(10) & 10.1\\
18 & 9 & 10 & 17 & 9 & 9 & 199,169.747(10) & 10.1\\
18 & 8 & 10 & 17 & 8 & 9 & 199,170.326(10) & -9.3\\
18 & 8 & 11 & 17 & 8 & 10 & 199,170.326(10) & -9.3\\
18 & 10 & 8 & 17 & 10 & 7 & 199,177.446(30) & -22.2\\
18 & 10 & 9 & 17 & 10 & 8 & 199,177.446(30) & -22.2\\
18 & 7 & 11 & 17 & 7 & 10 & 199,183.075(10) & -10.7\\
18 & 7 & 12 & 17 & 7 & 11 & 199,183.075(10) & -10.6\\
18 & 11 & 7 & 17 & 11 & 6 & 199,191.377(20) & -5.5\\
18 & 11 & 8 & 17 & 11 & 7 & 199,191.377(20) & -5.5\\
18 & 12 & 6 & 17 & 12 & 5 & 199,210.185(10) & -5.0\\
18 & 12 & 7 & 17 & 12 & 6 & 199,210.185(10) & -5.0\\
18 & 6 & 12 & 17 & 6 & 11 & 199,215.399(10) & -3.9\\
18 & 6 & 13 & 17 & 6 & 12 & 199,215.399(10) & 2.7\\
18 & 14 & 4 & 17 & 14 & 3 & 199,259.486(10) & 11.4\\
18 & 14 & 5 & 17 & 14 & 4 & 199,259.486(10) & 11.4\\
18 & 5 & 14 & 17 & 5 & 13 & 199,283.132(15) & 3.5\\
18 & 5 & 13 & 17 & 5 & 12 & 199,283.658(15) & -4.1\\
18 & 15 & 4 & 17 & 15 & 3 & 199,288.983(30) & -36.5\\
18 & 15 & 3 & 17 & 15 & 2 & 199,288.983(30) & -36.5\\
18 & 16 & 3 & 17 & 16 & 2 & 199,321.489(35) & 64.9\\
18 & 16 & 2 & 17 & 16 & 1 & 199,321.489(35) & 64.9\\
18 & 4 & 15 & 17 & 4 & 14 & 199,415.908(10) & -3.8\\
18 & 4 & 14 & 17 & 4 & 13 & 199,441.362(15) & -14.8\\
18 & 3 & 16 & 17 & 3 & 15 & 199,487.117(10) & 4.6\\
18 & 3 & 15 & 17 & 3 & 14 & 200,099.455(20) & 19.1\\
19 & 3 & 16 & 19 & 2 & 17 & 201,311.948(25) & -22.0\\
18 & 1 & 17 & 17 & 1 & 16 & 202,643.056(10) & -21.0\\
19 & 1 & 19 & 18 & 1 & 18 & 202,787.333(15) & -18.1\\
18 & 2 & 16 & 17 & 2 & 15 & 202,928.293(10) & -15.7\\
20 & 0 & 20 & 19 & 1 & 19 & 203,870.191(15) & -20.8\\
18 & 3 & 15 & 18 & 2 & 16 & 204,367.274(35) & -56.0\\
19 & 0 & 19 & 18 & 0 & 18 & 204,412.801(10) & -3.8\\
18 & 1 & 18 & 17 & 0 & 17 & 206,565.920(50) & 2.5\\
16 & 3 & 13 & 16 & 2 & 14 & 209,765.53(21) & 55.6\\
19 & 9 & 11 & 18 & 9 & 10 & 210,236.321(10) & 13.0\\
19 & 9 & 10 & 18 & 9 & 9 & 210,236.321(10) & 13.0\\
19 & 8 & 12 & 18 & 8 & 11 & 210,239.468(30) & 20.6\\
19 & 8 & 11 & 18 & 8 & 10 & 210,239.468(30) & 20.6\\
19 & 10 & 10 & 18 & 10 & 9 & 210,242.708(10) & 21.8\\
19 & 10 & 9 & 18 & 10 & 8 & 210,242.708(10) & 21.8\\
19 & 11 & 9 & 18 & 11 & 8 & 210,256.064(10) & 3.4\\
19 & 11 & 8 & 18 & 11 & 7 & 210,256.064(10) & 3.4\\
19 & 7 & 13 & 18 & 7 & 12 & 210,256.580(10) & -18.4\\
19 & 7 & 12 & 18 & 7 & 11 & 210,256.580(10) & -18.5\\
19 & 12 & 8 & 18 & 12 & 7 & 210,274.881(50) & -36.5\\
19 & 12 & 7 & 18 & 12 & 6 & 210,274.881(50) & -36.5\\
19 & 6 & 13 & 18 & 6 & 12 & 210,296.498(10) & 9.0\\
19 & 6 & 14 & 18 & 6 & 13 & 210,296.498(10) & 21.3\\
19 & 13 & 6 & 18 & 13 & 5 & 210,298.309(10) & 13.0\\
19 & 13 & 7 & 18 & 13 & 6 & 210,298.309(10) & 13.0\\
19 & 14 & 5 & 18 & 14 & 4 & 210,325.550(15) & -5.4\\
19 & 14 & 6 & 18 & 14 & 5 & 210,325.550(15) & -5.4\\
19 & 15 & 4 & 18 & 15 & 3 & 210,356.255(25) & 6.4\\
19 & 15 & 5 & 18 & 15 & 4 & 210,356.255(25) & 6.4\\
19 & 5 & 15 & 18 & 5 & 14 & 210,377.687(20) & 28.2\\
19 & 5 & 14 & 18 & 5 & 13 & 210,378.537(10) & 3.4\\
19 & 16 & 3 & 18 & 16 & 2 & 210,390.050(20) & -0.4\\
19 & 16 & 4 & 18 & 16 & 3 & 210,390.050(20) & -0.4\\
19 & 4 & 16 & 18 & 4 & 15 & 210,532.045(10) & 10.9\\
19 & 4 & 15 & 18 & 4 & 14 & 210,569.240(10) & -4.3\\
19 & 3 & 17 & 18 & 3 & 16 & 210,585.163(10) & 5.0\\
19 & 3 & 16 & 18 & 3 & 15 & 211,379.789(10) & 18.1\\
20 & 1 & 20 & 19 & 1 & 19 & 213,331.295(20) & 15.8\\
19 & 1 & 18 & 18 & 1 & 17 & 213,627.622(10) & 19.6\\
14 & 3 & 11 & 14 & 2 & 12 & 214,048.296(50) & 19.0\\
19 & 2 & 17 & 18 & 2 & 16 & 214,435.136(10) & 5.1\\
20 & 0 & 20 & 19 & 0 & 19 & 214,798.102(10) & 5.3\\
20 & 9 & 11 & 19 & 9 & 10 & 221,303.141(10) & 11.2\\
20 & 9 & 12 & 19 & 9 & 11 & 221,303.141(10) & 11.2\\
20 & 10 & 10 & 19 & 10 & 9 & 221,307.861(10) & -5.9\\
20 & 10 & 11 & 19 & 10 & 10 & 221,307.861(10) & -5.9\\
20 & 8 & 12 & 19 & 8 & 11 & 221,309.217(10) & 1.1\\
20 & 8 & 13 & 19 & 8 & 12 & 221,309.217(10) & 1.1\\
20 & 11 & 10 & 19 & 11 & 9 & 221,320.479(10) & -10.1\\
20 & 11 & 9 & 19 & 11 & 8 & 221,320.479(10) & -10.1\\
20 & 7 & 14 & 19 & 7 & 13 & 221,331.382(10) & 15.9\\
20 & 7 & 13 & 19 & 7 & 12 & 221,331.382(10) & 15.7\\
20 & 13 & 7 & 19 & 13 & 6 & 221,362.979(15) & -7.8\\
20 & 13 & 8 & 19 & 13 & 7 & 221,362.979(15) & -7.8\\
20 & 6 & 15 & 19 & 6 & 14 & 221,379.761(10) & 10.5\\
20 & 6 & 14 & 19 & 6 & 13 & 221,379.761(10) & -11.3\\
20 & 14 & 6 & 19 & 14 & 5 & 221,391.015(10) & 12.7\\
20 & 14 & 7 & 19 & 14 & 6 & 221,391.015(10) & 12.7\\
20 & 15 & 5 & 19 & 15 & 4 & 221,422.800(45) & 35.7\\
20 & 15 & 6 & 19 & 15 & 5 & 221,422.800(45) & 35.7\\
20 & 16 & 4 & 19 & 16 & 3 & 221,457.82(10) & -78.3\\
20 & 16 & 5 & 19 & 16 & 4 & 221,457.82(10) & -78.3\\
20 & 5 & 16 & 19 & 5 & 15 & 221,475.933(10) & 18.6\\
20 & 5 & 15 & 19 & 5 & 14 & 221,477.312(10) & 1.1\\
20 & 17 & 3 & 19 & 17 & 2 & 221,496.21(10) & 89.1\\
20 & 17 & 4 & 19 & 17 & 3 & 221,496.21(10) & 89.1\\
20 & 3 & 18 & 19 & 3 & 17 & 221,679.067(50) & 49.0\\
20 & 4 & 16 & 19 & 4 & 15 & 221,706.383(10) & 18.9\\
10 & 3 & 8 & 10 & 2 & 9 & 222,271.179(65) & -57.6\\
20 & 3 & 17 & 19 & 3 & 16 & 222,693.623(10) & -6.4\\
13 & 3 & 11 & 13 & 2 & 12 & 223,763.431(15) & 14.1\\
21 & 1 & 21 & 20 & 1 & 20 & 223,862.742(10) & -12.8\\
20 & 1 & 20 & 19 & 0 & 19 & 224,259.184(15) & 19.8\\
14 & 3 & 12 & 14 & 2 & 13 & 224,521.858(20) & 19.6\\
20 & 1 & 19 & 19 & 1 & 18 & 224,560.765(10) & -4.1\\
21 & 0 & 21 & 20 & 0 & 20 & 225,173.690(10) & -14.7\\
15 & 3 & 13 & 15 & 2 & 14 & 225,439.218(15) & 9.8\\
20 & 2 & 18 & 19 & 2 & 17 & 225,931.877(10) & 0.2\\
16 & 3 & 14 & 16 & 2 & 15 & 226,532.703(10) & 9.0\\
22 & 0 & 22 & 21 & 1 & 21 & 227,393.803(15) & -17.0\\
17 & 3 & 15 & 17 & 2 & 16 & 227,819.232(20) & -27.2\\
18 & 3 & 16 & 18 & 2 & 17 & 229,315.465(10) & -8.8\\
21 & 2 & 20 & 20 & 2 & 19 & 230,549.163(10) & -20.9\\
21 & 9 & 12 & 20 & 9 & 11 & 232,370.207(10) & -6.8\\
21 & 9 & 13 & 20 & 9 & 12 & 232,370.207(10) & -6.8\\
21 & 10 & 11 & 20 & 10 & 10 & 232,373.002(10) & -4.7\\
21 & 10 & 12 & 20 & 10 & 11 & 232,373.002(10) & -4.7\\
21 & 8 & 13 & 20 & 8 & 12 & 232,379.674(10) & 0.3\\
21 & 8 & 14 & 20 & 8 & 13 & 232,379.674(10) & 0.3\\
21 & 11 & 10 & 20 & 11 & 9 & 232,384.630(10) & -23.8\\
21 & 11 & 11 & 20 & 11 & 10 & 232,384.630(10) & -23.8\\
21 & 12 & 9 & 20 & 12 & 8 & 232,403.115(10) & -4.0\\
21 & 12 & 10 & 20 & 12 & 9 & 232,403.115(10) & -4.0\\
21 & 7 & 15 & 20 & 7 & 14 & 232,407.448(10) & -5.4\\
21 & 7 & 14 & 20 & 7 & 13 & 232,407.448(10) & -5.9\\
21 & 13 & 8 & 20 & 13 & 7 & 232,427.083(50) & -30.6\\
21 & 13 & 9 & 20 & 13 & 8 & 232,427.083(50) & -30.6\\
21 & 14 & 7 & 20 & 14 & 6 & 232,455.767(10) & -14.3\\
21 & 14 & 8 & 20 & 14 & 7 & 232,455.767(10) & -14.3\\
21 & 6 & 15 & 20 & 6 & 14 & 232,465.347(10) & -23.4\\
21 & 6 & 16 & 20 & 6 & 15 & 232,465.347(10) & 14.3\\
21 & 15 & 7 & 20 & 15 & 6 & 232,488.509(25) & -19.4\\
21 & 15 & 6 & 20 & 15 & 5 & 232,488.509(25) & -19.4\\
21 & 16 & 6 & 20 & 16 & 5 & 232,524.921(10) & -5.2\\
21 & 16 & 5 & 20 & 16 & 4 & 232,524.921(10) & -5.2\\
21 & 17 & 5 & 20 & 17 & 4 & 232,564.674(45) & 21.1\\
21 & 17 & 4 & 20 & 17 & 3 & 232,564.674(45) & 21.1\\
21 & 5 & 17 & 20 & 5 & 16 & 232,578.069(10) & 0.3\\
21 & 5 & 16 & 20 & 5 & 15 & 232,580.247(10) & 2.0\\
21 & 3 & 19 & 20 & 3 & 18 & 232,767.194(10) & 6.7\\
21 & 4 & 18 & 20 & 4 & 17 & 232,778.976(10) & 4.2\\
21 & 4 & 17 & 20 & 4 & 16 & 232,853.815(10) & 22.2\\
21 & 3 & 18 & 20 & 3 & 17 & 234,043.345(10) & 8.9\\
22 & 0 & 22 & 21 & 0 & 21 & 235,543.961(10) & 23.3\\
21 & 2 & 19 & 20 & 2 & 18 & 237,411.402(10) & -16.9\\
23 & 0 & 23 & 22 & 1 & 22 & 238,923.506(10) & 10.0\\
22 & 2 & 21 & 21 & 2 & 20 & 241,361.566(10) & 8.7\\
22 & 1 & 22 & 21 & 0 & 21 & 242,532.596(55) & 12.1\\
22 & 9 & 14 & 21 & 9 & 13 & 243,437.575(15) & 3.5\\
22 & 9 & 13 & 21 & 9 & 12 & 243,437.575(15) & 3.5\\
22 & 10 & 13 & 21 & 10 & 12 & 243,438.097(45) & -5.2\\
22 & 10 & 12 & 21 & 10 & 11 & 243,438.097(45) & -5.2\\
22 & 11 & 12 & 21 & 11 & 11 & 243,448.530(10) & -10.1\\
22 & 11 & 11 & 21 & 11 & 10 & 243,448.530(10) & -10.1\\
22 & 8 & 14 & 21 & 8 & 13 & 243,450.900(50) & 46.6\\
22 & 8 & 15 & 21 & 8 & 14 & 243,450.900(50) & 46.6\\
22 & 12 & 10 & 21 & 12 & 9 & 243,466.574(15) & 26.3\\
22 & 12 & 11 & 21 & 12 & 10 & 243,466.574(15) & 26.3\\
22 & 7 & 16 & 21 & 7 & 15 & 243,484.952(15) & 27.0\\
22 & 7 & 15 & 21 & 7 & 14 & 243,484.952(15) & 26.1\\
22 & 14 & 8 & 21 & 14 & 7 & 243,519.828(50) & -30.1\\
22 & 14 & 9 & 21 & 14 & 8 & 243,519.828(50) & -30.1\\
22 & 6 & 16 & 21 & 6 & 15 & 243,553.415(50) & 12.9\\
22 & 6 & 17 & 21 & 6 & 16 & 243,553.415(50) & 76.5\\
22 & 5 & 18 & 21 & 5 & 17 & 243,684.282(10) & -4.0\\
22 & 5 & 17 & 21 & 5 & 16 & 243,687.572(50) & -32.8\\
22 & 3 & 20 & 21 & 3 & 19 & 243,848.117(10) & 4.7\\
22 & 4 & 19 & 21 & 4 & 18 & 243,909.394(10) & -1.7\\
22 & 4 & 18 & 21 & 4 & 17 & 244,012.745(15) & 17.2\\
23 & 1 & 23 & 22 & 1 & 22 & 244,891.145(10) & 3.8\\
22 & 3 & 19 & 21 & 3 & 18 & 245,430.646(10) & -11.9\\
23 & 0 & 23 & 22 & 0 & 22 & 245,912.148(10) & 5.8\\
22 & 1 & 21 & 21 & 1 & 20 & 246,260.384(10) & 10.4\\
22 & 2 & 20 & 21 & 2 & 19 & 248,867.410(40) & -19.5\\
\end{supertabular}

\newpage

\tablehead{
\toprule
 $J'$ & $K'_a$ & $K'_c$ & $J''$ & $K''_a$ & $K''_c$ & \multicolumn{1}{c}{Obs.} & \multicolumn{1}{c}{Obs - Calc.} \\
\midrule}
\tabletail{\bottomrule}
\topcaption{Observed frequencies for \textit{anti}-vinyl mercaptan. Frequencies are given in MHz. Values in parentheses represent $1\sigma$ uncertainty. The obs. -- calc. values are given in kHz.}
\begin{supertabular}{c c c c c c . .}
1 & 0 & 1 & 0 & 0 & 0 & 11,176.6389(10) & 0.2\\
5 & 0 & 5 & 4 & 1 & 4 & 14,923.9484(10) & -0.3\\
1 & 1 & 1 & 2 & 0 & 2 & 21,178.9543(10) & -0.4\\
2 & 1 & 2 & 1 & 1 & 1 & 21,735.5955(10) & -1.1\\
2 & 0 & 2 & 1 & 0 & 1 & 22,346.6731(10) & -0.4\\
2 & 1 & 1 & 1 & 1 & 0 & 22,971.1109(10) & 0.3\\
6 & 0 & 6 & 5 & 1 & 5 & 27,443.7460(10) & -0.2\\
3 & 1 & 3 & 2 & 1 & 2 & 32,599.1768(10) & -0.7\\
3 & 0 & 3 & 2 & 0 & 2 & 33,503.5068(10) & 0.4\\
3 & 1 & 2 & 2 & 1 & 1 & 34,452.3503(20) & 1.8\\
12 & 1 & 11 & 12 & 0 & 12 & 72,570.263(30) & -73.7\\
7 & 1 & 7 & 6 & 1 & 6 & 75,988.301(15) & 6.6\\
7 & 2 & 6 & 6 & 2 & 5 & 78,183.738(20) & 51.4\\
7 & 2 & 5 & 6 & 2 & 4 & 78,546.705(40) & 13.1\\
10 & 0 & 10 & 9 & 1 & 9 & 79,116.675(10) & -4.3\\
7 & 1 & 6 & 6 & 1 & 5 & 80,305.997(10) & 19.6\\
4 & 1 & 4 & 3 & 0 & 3 & 85,467.985(10) & 4.0\\
8 & 1 & 8 & 7 & 1 & 7 & 86,812.171(10) & -0.8\\
8 & 0 & 8 & 7 & 0 & 7 & 88,864.736(10) & -6.1\\
8 & 2 & 7 & 7 & 2 & 6 & 89,329.569(20) & -12.5\\
8 & 4 & 4 & 7 & 4 & 3 & 89,462.308(10) & -32.5\\
8 & 4 & 5 & 7 & 4 & 4 & 89,462.308(10) & 45.3\\
8 & 3 & 6 & 7 & 3 & 5 & 89,485.192(10) & -21.6\\
8 & 3 & 5 & 7 & 3 & 4 & 89,496.052(25) & 1.9\\
8 & 2 & 6 & 7 & 2 & 5 & 89,871.505(25) & 3.0\\
8 & 1 & 7 & 7 & 1 & 6 & 91,741.577(10) & 3.4\\
11 & 0 & 11 & 10 & 1 & 10 & 92,238.816(10) & -3.3\\
5 & 1 & 5 & 4 & 0 & 4 & 95,137.182(10) & -6.2\\
9 & 1 & 9 & 8 & 1 & 8 & 97,624.268(10) & 4.0\\
9 & 0 & 9 & 8 & 0 & 8 & 99,811.519(15) & -11.5\\
9 & 2 & 8 & 8 & 2 & 7 & 100,466.233(10) & 11.2\\
9 & 6 & 4 & 8 & 6 & 3 & 100,637.563(10) & 14.5\\
9 & 6 & 3 & 8 & 6 & 2 & 100,637.563(10) & 14.5\\
9 & 5 & 5 & 8 & 5 & 4 & 100,640.558(50) & 35.5\\
9 & 5 & 4 & 8 & 5 & 3 & 100,640.558(50) & 34.6\\
9 & 4 & 5 & 8 & 4 & 4 & 100,654.312(50) & -57.8\\
9 & 4 & 6 & 8 & 4 & 5 & 100,654.312(50) & 128.8\\
9 & 3 & 7 & 8 & 3 & 6 & 100,685.985(10) & -7.4\\
9 & 3 & 6 & 8 & 3 & 5 & 100,705.840(10) & 5.8\\
9 & 2 & 7 & 8 & 2 & 6 & 101,235.165(10) & 10.9\\
9 & 1 & 8 & 8 & 1 & 7 & 103,161.608(10) & -2.3\\
6 & 1 & 6 & 5 & 0 & 5 & 104,539.564(10) & -6.1\\
12 & 0 & 12 & 11 & 1 & 11 & 105,359.766(10) & 9.9\\
10 & 0 & 10 & 9 & 0 & 9 & 110,705.947(10) & 3.0\\
10 & 2 & 9 & 9 & 2 & 8 & 111,592.467(10) & -6.9\\
10 & 6 & 4 & 9 & 6 & 3 & 111,823.033(10) & 1.6\\
10 & 6 & 5 & 9 & 6 & 4 & 111,823.033(10) & 1.6\\
10 & 5 & 5 & 9 & 5 & 4 & 111,829.035(10) & -7.2\\
10 & 5 & 6 & 9 & 5 & 5 & 111,829.035(10) & -4.9\\
10 & 4 & 7 & 9 & 4 & 6 & 111,849.294(10) & 1.8\\
10 & 4 & 6 & 9 & 4 & 5 & 111,849.693(10) & -3.2\\
10 & 3 & 8 & 9 & 3 & 7 & 111,890.948(10) & 0.7\\
10 & 2 & 8 & 9 & 2 & 7 & 112,639.999(10) & -2.9\\
7 & 1 & 7 & 6 & 0 & 6 & 113,698.287(10) & -5.8\\
10 & 1 & 9 & 9 & 1 & 8 & 114,563.588(10) & 1.0\\
13 & 0 & 13 & 12 & 1 & 12 & 118,439.637(10) & -0.4\\
11 & 1 & 11 & 10 & 1 & 10 & 119,209.414(10) & -5.8\\
11 & 1 & 11 & 10 & 1 & 10 & 119,209.414(10) & -5.8\\
11 & 0 & 11 & 10 & 0 & 10 & 121,545.752(20) & 0.3\\
8 & 1 & 8 & 7 & 0 & 7 & 122,641.283(20) & 9.7\\
11 & 2 & 10 & 10 & 2 & 9 & 122,707.191(45) & -27.3\\
16 & 0 & 16 & 15 & 0 & 15 & 174,966.024(30) & -8.0\\
16 & 8 & 9 & 15 & 8 & 8 & 178,940.02(12) & -10.6\\
16 & 8 & 8 & 15 & 8 & 7 & 178,940.02(12) & -10.6\\
16 & 7 & 10 & 15 & 7 & 9 & 178,944.269(10) & 23.8\\
16 & 7 & 9 & 15 & 7 & 8 & 178,944.269(10) & 23.8\\
16 & 9 & 7 & 15 & 9 & 6 & 178,945.233(40) & 45.4\\
16 & 9 & 8 & 15 & 9 & 7 & 178,945.233(40) & 45.4\\
16 & 10 & 7 & 15 & 10 & 6 & 178,956.988(15) & 36.9\\
16 & 10 & 6 & 15 & 10 & 5 & 178,956.988(15) & 36.9\\
16 & 6 & 10 & 15 & 6 & 9 & 178,963.196(25) & 36.1\\
16 & 6 & 11 & 15 & 6 & 10 & 178,963.196(25) & 38.0\\
16 & 11 & 6 & 15 & 11 & 5 & 178,973.765(40) & -1.1\\
16 & 11 & 5 & 15 & 11 & 4 & 178,973.765(40) & -1.1\\
16 & 12 & 5 & 15 & 12 & 4 & 178,994.759(55) & 68.6\\
16 & 12 & 4 & 15 & 12 & 3 & 178,994.759(55) & 68.6\\
16 & 5 & 12 & 15 & 5 & 11 & 179,008.358(30) & 119.1\\
16 & 5 & 11 & 15 & 5 & 10 & 179,008.358(30) & -75.4\\
16 & 4 & 13 & 15 & 4 & 12 & 179,103.592(60) & 61.2\\
16 & 4 & 12 & 15 & 4 & 11 & 179,115.290(60) & 14.2\\
16 & 3 & 14 & 15 & 3 & 13 & 179,187.710(55) & -95.6\\
16 & 3 & 13 & 15 & 3 & 12 & 179,546.12(17) & 219.0\\
16 & 2 & 14 & 15 & 2 & 13 & 181,867.564(30) & -39.2\\
16 & 1 & 15 & 15 & 1 & 14 & 182,430.250(30) & -8.0\\
17 & 0 & 17 & 16 & 0 & 16 & 185,524.475(15) & -7.0\\
17 & 8 & 9 & 16 & 8 & 8 & 190,126.466(20) & 23.4\\
17 & 8 & 10 & 16 & 8 & 9 & 190,126.466(20) & 23.4\\
17 & 9 & 9 & 16 & 9 & 8 & 190,129.859(10) & -11.1\\
17 & 9 & 8 & 16 & 9 & 7 & 190,129.859(10) & -11.1\\
17 & 7 & 10 & 16 & 7 & 9 & 190,133.918(15) & -16.9\\
17 & 7 & 11 & 16 & 7 & 10 & 190,133.918(15) & -16.9\\
17 & 10 & 7 & 16 & 10 & 6 & 190,140.914(40) & 2.4\\
17 & 10 & 8 & 16 & 10 & 7 & 190,140.914(40) & 2.4\\
17 & 11 & 7 & 16 & 11 & 6 & 190,157.712(25) & 8.0\\
17 & 11 & 6 & 16 & 11 & 5 & 190,157.712(25) & 8.0\\
17 & 6 & 11 & 16 & 6 & 10 & 190,158.762(20) & 15.7\\
17 & 6 & 12 & 16 & 6 & 11 & 190,158.762(20) & 19.5\\
17 & 5 & 12 & 16 & 5 & 11 & 190,214.765(15) & -162.1\\
17 & 5 & 13 & 16 & 5 & 12 & 190,214.765(15) & 177.6\\
17 & 4 & 14 & 16 & 4 & 13 & 190,328.673(25) & 14.7\\
17 & 4 & 13 & 16 & 4 & 12 & 190,346.662(20) & -13.7\\
17 & 3 & 15 & 16 & 3 & 14 & 190,408.156(25) & 18.2\\
17 & 3 & 14 & 16 & 3 & 13 & 190,890.135(45) & -8.6\\
17 & 2 & 15 & 16 & 2 & 14 & 193,492.836(15) & 11.1\\
17 & 1 & 16 & 16 & 1 & 15 & 193,618.481(25) & 24.0\\
18 & 0 & 18 & 17 & 0 & 17 & 196,055.843(15) & 19.3\\
18 & 2 & 17 & 17 & 2 & 16 & 200,099.455(15) & -29.1\\
18 & 8 & 11 & 17 & 8 & 10 & 201,313.343(60) & 13.8\\
18 & 8 & 10 & 17 & 8 & 9 & 201,313.343(60) & 13.8\\
18 & 9 & 9 & 17 & 9 & 8 & 201,314.711(30) & 53.3\\
18 & 9 & 10 & 17 & 9 & 9 & 201,314.711(30) & 53.3\\
18 & 11 & 8 & 17 & 11 & 7 & 201,341.278(40) & -10.5\\
18 & 11 & 7 & 17 & 11 & 6 & 201,341.278(40) & -10.5\\
18 & 6 & 12 & 17 & 6 & 11 & 201,356.236(15) & 19.8\\
18 & 6 & 13 & 17 & 6 & 12 & 201,356.236(15) & 27.1\\
18 & 12 & 7 & 17 & 12 & 6 & 201,363.081(60) & 27.8\\
18 & 12 & 6 & 17 & 12 & 5 & 201,363.081(60) & 27.8\\
18 & 13 & 6 & 17 & 13 & 5 & 201,389.094(50) & -61.9\\
18 & 13 & 5 & 17 & 13 & 4 & 201,389.094(50) & -61.9\\
18 & 5 & 14 & 17 & 5 & 13 & 201,424.312(50) & 68.0\\
18 & 5 & 13 & 17 & 5 & 12 & 201,424.96(15) & 143.2\\
18 & 3 & 16 & 17 & 3 & 15 & 201,626.716(10) & -8.8\\
18 & 3 & 15 & 18 & 2 & 16 & 201,827.739(50) & -52.4\\
18 & 3 & 15 & 17 & 3 & 14 & 202,263.080(30) & 6.9\\
18 & 1 & 17 & 17 & 1 & 16 & 204,761.414(70) & 9.7\\
19 & 0 & 19 & 18 & 0 & 18 & 206,566.727(15) & -11.1\\
18 & 1 & 18 & 17 & 0 & 17 & 208,219.064(20) & -2.8\\
19 & 9 & 11 & 18 & 9 & 10 & 212,499.557(10) & 2.1\\
19 & 9 & 10 & 18 & 9 & 9 & 212,499.557(10) & 2.1\\
19 & 8 & 11 & 18 & 8 & 10 & 212,500.672(50) & -46.8\\
19 & 8 & 12 & 18 & 8 & 11 & 212,500.672(50) & -46.8\\
19 & 10 & 9 & 18 & 10 & 8 & 212,508.372(35) & 27.1\\
19 & 10 & 10 & 18 & 10 & 9 & 212,508.372(35) & 27.1\\
19 & 11 & 8 & 18 & 11 & 7 & 212,524.496(15) & -1.6\\
19 & 11 & 9 & 18 & 11 & 8 & 212,524.496(15) & -1.6\\
19 & 12 & 7 & 18 & 12 & 6 & 212,546.370(30) & -82.8\\
19 & 12 & 8 & 18 & 12 & 7 & 212,546.370(30) & -82.8\\
19 & 6 & 14 & 18 & 6 & 13 & 212,555.670(10) & 3.1\\
19 & 6 & 13 & 18 & 6 & 12 & 212,555.670(10) & -10.3\\
19 & 13 & 6 & 18 & 13 & 5 & 212,573.212(70) & -2.9\\
19 & 13 & 7 & 18 & 13 & 6 & 212,573.212(70) & -2.9\\
19 & 14 & 6 & 18 & 14 & 5 & 212,604.012(60) & -102.9\\
19 & 14 & 5 & 18 & 14 & 4 & 212,604.012(60) & -102.9\\
19 & 5 & 15 & 18 & 5 & 14 & 212,637.387(60) & -3.2\\
19 & 5 & 14 & 18 & 5 & 13 & 212,638.355(15) & 24.1\\
19 & 4 & 15 & 18 & 4 & 14 & 212,833.069(20) & -38.8\\
20 & 1 & 20 & 19 & 1 & 19 & 215,625.949(10) & 9.1\\
20 & 0 & 20 & 19 & 0 & 19 & 217,063.251(50) & -50.6\\
7 & 3 & 4 & 7 & 2 & 5 & 218,549.258(10) & 2.0\\
20 & 2 & 19 & 19 & 2 & 18 & 222,051.307(25) & 26.1\\
20 & 9 & 12 & 19 & 9 & 11 & 223,684.563(10) & -3.2\\
20 & 9 & 11 & 19 & 9 & 10 & 223,684.563(10) & -3.2\\
20 & 8 & 12 & 19 & 8 & 11 & 223,688.610(15) & -27.6\\
20 & 8 & 13 & 19 & 8 & 12 & 223,688.610(15) & -27.6\\
20 & 10 & 11 & 19 & 10 & 10 & 223,691.759(20) & -37.2\\
20 & 10 & 10 & 19 & 10 & 9 & 223,691.759(20) & -37.2\\
20 & 11 & 9 & 19 & 11 & 8 & 223,707.337(20) & 27.7\\
20 & 11 & 10 & 19 & 11 & 9 & 223,707.337(20) & 27.7\\
20 & 7 & 14 & 19 & 7 & 13 & 223,709.360(10) & -24.3\\
20 & 7 & 13 & 19 & 7 & 12 & 223,709.360(10) & -24.5\\
20 & 12 & 8 & 19 & 12 & 7 & 223,729.335(20) & 44.3\\
20 & 12 & 9 & 19 & 12 & 8 & 223,729.335(20) & 44.3\\
20 & 13 & 7 & 19 & 13 & 6 & 223,756.47(13) & -112.6\\
20 & 13 & 8 & 19 & 13 & 7 & 223,756.47(13) & -112.6\\
20 & 6 & 15 & 19 & 6 & 14 & 223,757.222(10) & -4.3\\
20 & 6 & 14 & 19 & 6 & 13 & 223,757.222(10) & -28.1\\
20 & 5 & 16 & 19 & 5 & 15 & 223,854.290(40) & 88.4\\
20 & 5 & 15 & 19 & 5 & 14 & 223,855.66(16) & -46.1\\
20 & 4 & 17 & 19 & 4 & 16 & 224,033.610(20) & -24.4\\
20 & 3 & 18 & 19 & 3 & 17 & 224,053.046(70) & 59.1\\
20 & 4 & 16 & 19 & 4 & 15 & 224,089.780(65) & -169.7\\
20 & 3 & 17 & 19 & 3 & 16 & 225,106.370(10) & 1.3\\
20 & 1 & 19 & 19 & 1 & 18 & 226,896.237(10) & 25.5\\
\end{supertabular}

\end{document}